\documentclass[aps,twocolumn,prb,superscriptaddress]{revtex4-2}
\usepackage{graphicx}
\usepackage[utf8]{inputenc}
\usepackage{amsmath,amssymb}
\usepackage{newtxtext}
\usepackage[varvw]{newtxmath}
\usepackage{mathtools}
\usepackage[pdftex]{hyperref}

\hypersetup{
    colorlinks,
    linkcolor={blue},
    citecolor={blue},
    urlcolor={blue}
}

\def\<{\langle}
\def\>{\rangle}

\DeclareMathOperator{\sgn}{sgn}

\renewcommand{\Im}[0]{\text{Im}\,}
\renewcommand{\Re}[0]{\text{Re}\,}

\newcommand{\ve}[1]{\boldsymbol{#1}}
\newcommand{\vmf}[0]{v_{\text{MF}}}
\newcommand{\mf}{\text{MF}}

\newcommand{\Jk}{J_{\mathrm{K}}}
\newcommand{\Jkc}{J_{\mathrm{K}}^{\mathrm{c}}}
\newcommand{\Jh}{J_{\mathrm{H}}}
\newcommand{\geff}{g_{\text{eff}}}

\makeatletter
\def\maketitle{
\@author@finish
\title@column\titleblock@produce
\suppressfloats[t]}

\makeatother

\makeindex

\date{\today}

\begin{document}

\title{
Marginal Fermi liquid at magnetic quantum criticality from dimensional confinement
}

\author{Bernhard Frank}
\affiliation{Institut f\"ur Theoretische Physik and W\"urzburg-Dresden Cluster of Excellence ct.qmat,
Technische Universit\"at Dresden, 01062 Dresden, Germany}
\author{Zi Hong Liu}
\affiliation{Institut f\"ur Theoretische Physik und Astrophysik and W\"urzburg-Dresden Cluster of Excellence ct.qmat,\\
Universit\"at W\"urzburg, 97074 W\"urzburg, Germany}
\author{Fakher F. Assaad}
\affiliation{Institut f\"ur Theoretische Physik und Astrophysik and W\"urzburg-Dresden Cluster of Excellence ct.qmat,\\
Universit\"at W\"urzburg, 97074 W\"urzburg, Germany}
\author{Matthias Vojta}
\affiliation{Institut f\"ur Theoretische Physik and W\"urzburg-Dresden Cluster of Excellence ct.qmat,
Technische Universit\"at Dresden, 01062 Dresden, Germany}
\author{Lukas Janssen}
\affiliation{Institut f\"ur Theoretische Physik and W\"urzburg-Dresden Cluster of Excellence ct.qmat,
Technische Universit\"at Dresden, 01062 Dresden, Germany}

\begin{abstract}
Metallic quantum criticality is frequently discussed as a source for non-Fermi liquid behavior, but controlled theoretical treatments are scarce. Here we identify and study a novel magnetic quantum critical point in a two-dimensional antiferromagnet coupled to a three-dimensional environment of conduction electrons. Using sign-problem-free quantum Monte Carlo simulations and an effective field-theory analysis, we demonstrate that the quantum critical point is characterized by marginal Fermi liquid behavior. In particular, we compute the electrical resistivity for transport across the magnetic layer, which is shown to display a linear temperature dependence at criticality. Experimental realizations in Kondo heterostructures are discussed.
\end{abstract}

\maketitle

%%%%%%%%%%%%%%%%%%%%%%%%%%%%%%%%%%%%%%%%%%%%%%%%%%%%%%%%%%%%%%%%%%%%%%%

Non-Fermi liquid behavior, including that of strange metals, is often observed in correlated metals, but despite intense recent research \cite{Stewart_RMP,lee2018recent,hussey22}, its origins remain poorly understood. Conceptually, one needs to distinguish cases where a description in terms of electronic quasiparticles remains valid, with potentially singular corrections to observables arising from their scattering, from those where electronic quasiparticles cease to be well-defined. The boundary between the two has been dubbed marginal Fermi liquid, characterized by a linear-in-energy scattering rate \cite{varma1989,varma_physrep}.
Deviations from Fermi-liquid behavior may occur in stable phases of matter, driven by either strong interactions, quenched disorder, or a combination of both, or it may originate from zero-temperature transitions between different quantum phases and the associated fluctuations \cite{sachdev_book,2007_HvL_RMP,gegenwart2008quantum}. 
The latter, referred to as metallic quantum criticality,  has been studied experimentally in a variety of heavy-fermion metals \cite{wirth16}, ${}^3$He bilayers \cite{neumann07}, and, most recently, in MoTe$_2$/WSe$_2$ moir\'e heterostructures \cite{zhao22}.
Theoretical investigations date back to the work of Hertz \cite{hertz1976}, Millis \cite{millis1993}, and Moriya \cite{moriya2012spin}, who used a perturbative framework in the spirit of Landau-Ginzburg-Wilson (LGW) to capture the physics of order-parameter fluctuations and their effects on electronic properties. Subsequent work showed that the LGW treatment is insufficient in two space dimensions, and more refined theories have been considered \cite{abanov2004anomalous,schlief2017exact,lee2018recent}. Their predictions have been compared to the results of extensive quantum Monte Carlo (QMC) simulations, with partial success \cite{berg_rev}. A common problem of many simulations is that it is difficult to reach sufficiently low temperatures to access the asymptotic quantum critical regime.
In this paper, we identify an example for a metallic quantum phase transition that is both analytically and computationally tractable, and realizes a novel instance of a marginal Fermi liquid. This is achieved by dimensional mismatch \cite{Danu20,Danu22}: We consider a Kondo-lattice-type model \cite{danu21} describing a two-dimensional (2D) local-moment magnet embedded in a three-dimensional (3D) metal. Such a setting has been proposed in Ref.~\cite{liu2022} and may be experimentally realized in heterostructures of layered materials, such as a single CeIn$_3$ layer embedded in bulk LaIn$_3$ \cite{shishido10}. 
%
%\todo{ref to recent Moire Kondo ideas? however, this is never 3D}  \fa{You  are referring  to  \href{https://arxiv.org/pdf/2211.00263.pdf} {this}. It  is  certainly   very  interesting  and  reminds  me on the  works of   \href{https://www.science.org/doi/abs/10.1126/science.1143607}{Saunders}   Both  are  2D.  } 
%\lj{Shifted this to first paragraph}
%
Importantly, the Kondo interaction between the two subsystems suppresses magnetic order and hence enables one to tune the 2D magnet to a quantum critical point. At this critical point, the heavy quasiparticles acquire a marginal Fermi-liquid self-energy, and the electrical resistivity measured across the magnetic layer has a linear temperature dependence at low $T$, see Fig.~\ref{fig:SysTrans}.

\begin{figure}[b]
\includegraphics[width=\columnwidth]{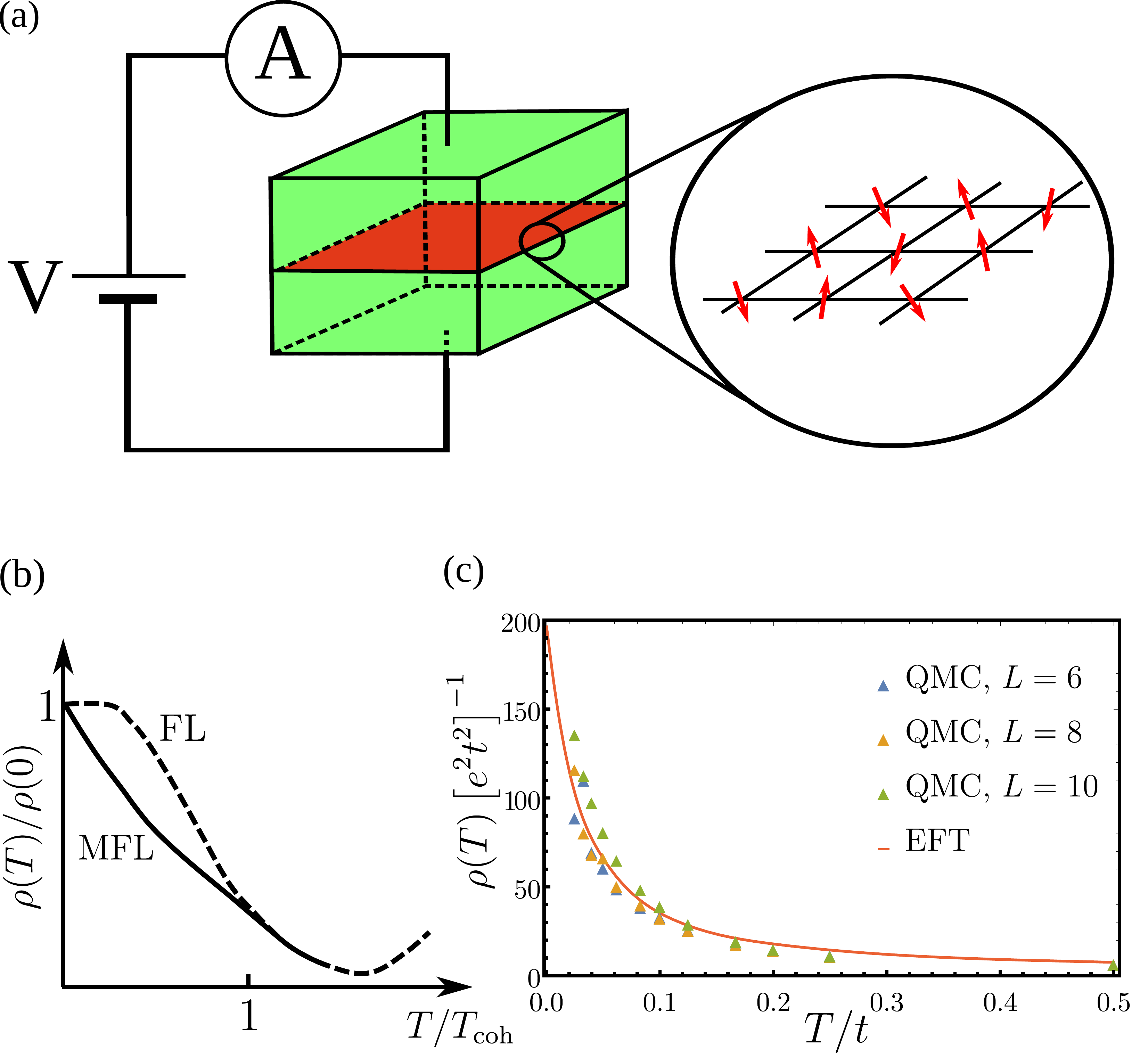}
\caption{%
(a) Schematic setup of proposed transport experiment on Kondo heterostructure.
(b) Schematic temperature dependence of resistivity $\rho(T)$: The marginal Fermi liquid at the quantum critical point gives rise to a linear decrease with temperature for temperatures below the coherence scale $T_\text{coh}$, in contrast to the conventional quadratic decrease in case of a Fermi liquid. At larger temperatures beyond $T_\text{coh}$ intrinsic interaction effects in the metal lead to an increase of $\rho(T)$ (dashed).
(c) Resistivity at the quantum critical point from effective field theory and QMC simulations, both indicating the expected linear scaling, and very good quantitative agreement.
%
%\lj{Show only $L=10$ ($L=12$?)?}
%\lj{Show absolute units in (c), to emphasize quantitative agreement?}
%
%\lj{Show resistance instead of resistivity?}
%\lj{Decrease labels ``(a)'', ``(b)'', ``(c)'' (to about the same as caption font?)}
%\lj{Increase axes labels in (c)}
}
\label{fig:SysTrans}
\end{figure}

%%%%%%%%%%%%%%%%%%%%%%%%%%%%%%%%%%%%%%%%%%%%%%%%%%%%%%%%%%%%%%%%%%%%%%%

\paragraph*{Model.}
We model the Kondo heterostructure, Fig.~\ref{fig:SysTrans}(a),
by the Hamiltonian \cite{liu2022}
\begin{align}\label{eq:model}
H = \sum_{\ve k,\sigma} \epsilon_{\ve k} c^\dagger_{\ve k \sigma} c^{}_{\ve k \sigma} + \Jh \sum_{\left\< \ve i,\ve j \right \>} \hat{\ve S}^f_{\ve i} \cdot \hat{\ve S}^f_{\ve j} + \Jk \sum_{\ve i} \hat{\ve S}^c_{\ve i,R_z=0} \cdot \hat{\ve S}^f_{\ve i} \, ,
\end{align}
with 3D conduction-electron dispersion $\epsilon_{\ve k} = -2 t (\cos k_x +\cos k_y + \cos k_z)$, at half filling. $\hat{\ve S}^f$ describes local spin-1/2 degrees of freedom that reside on the sites of the square lattice $\ve i$ in the layer at $R_z=0$, and interact via a fixed nearest-neighbor Heisenberg coupling $\Jh=t/2$.  
The last term parametrized by $\Jk$ describes the Kondo interaction between the local moments and the spin density of the conduction electrons $\hat{\ve S}^c_{\ve i,R_z=0}= 1/2 \sum_{\sigma,\sigma'} c^\dagger_{\ve i,R_z=0,\sigma} \boldsymbol{\sigma}_{\sigma \sigma'} c_{\ve i,R_z=0,\sigma'}$ within the layer $R_z=0$. The presence of the local moments breaks momentum conservation in the out-of-plane direction, such that $k_z$ is not a good quantum number, whereas the in-plane momentum $\ve k_2=(k_x,k_y)$ is conserved.

The model has been studied previously using sign-problem-free QMC simulations~\cite{liu2022,ALF_v2}.
It forms an antiferromagnetic (AFM) heavy-fermion metal for small $\Jk$, and undergoes a continuous quantum phase transition to a paramagnetic heavy-fermion metal at the critical coupling $\Jkc/t = 3.019(4)$.
In particular, the dimensional mismatch gives rise to a metallic ground state despite the presence of particle-hole symmetry, which entails Kondo insulators in spatially homogeneous systems.
%The metallic character of the ground state result from the dimensional mismatch between the electronic and the local moment subsystems and is remarkable because the particle-hole symmetry of $H$ would entail insulating states in spatially homogeneous systems.
%
The absence of Kondo breakdown is indicated by a finite spectral weight of the composite fermion~\cite{danu21} 
$\psi^\dagger_{\ve i,\sigma} = \sum_{\sigma'} c^\dagger_{\ve i,R_z=0,\sigma'} \boldsymbol \sigma_{\sigma \sigma'}  \cdot \hat{\ve S}^f_{\ve i}$
at small energies below the  coherence  scale $T_\text{coh}$ throughout the phase diagram.
%, irrespective of $\Jk$.  
Therefore, Ref.~\cite{liu2022} concluded that the quantum critical point (QCP) 
can be described in a Landau-Ginzburg-Wilson setup, with a dynamical exponent $z=2$.
%while 
%the evaluation of the simulated correlation functions comply with the dynamical critical exponent $z=2$. 
In the following, we set up an effective field theory to analyze the effect of the critical fluctuations on the fermion excitations, which are shown to exhibit marginal Fermi liquid behavior~\cite{2007_HvL_RMP}.

%Like in Kondo lattices the system forms a Néel-ordered antiferromagent at small couplings and undergoes a continuous quantum phase transition at the critical coupling strength $(J_K)_c=3.04$ beyond which magnetic order ceases to exist. On the other hand, the presence of Kondo singlets leads to hybridization between the conduction electrons and the local moments as in heavy Fermion materials. Within QMC simulations these effects can be accessed by studying the composite Fermion operators
%\begin{align}\label{eq:defpsi}
%\psi^\dagger_{\ve R_2,\sigma} = \sum_{\sigma'} c^\dagger_{\ve R_2,R_z=0,\sigma'} \boldsymbol \sigma_{\sigma \sigma'}  \cdot \hat{\ve S}^f_{\ve R_2} \, ,
%\end{align}
%which are limited to the spin layer in case of the heterostructure. QMC simulations observe a finite spectral weight of $\psi_\sigma$ both in the symmetric and in the antiferromagnetic phase which implies the absence of Kondo breakdown. At $(J_K)_c$ hybridization remains strong such that the major part of the spectral weight is carried by the $\psi$ Fermions. Moreover, the fact that the conduction electrons explore the out-of-plane dimension gives rise to a finite density of states $g_{\psi}(\omega)$ at small frequencies which implies metallic behavior in marked contrast to homogoneous Kondo lattices with half-filled conduction bands that turn into Kondo insulators.

%%%%%%%%%%%%%%%%%%%%%%%%%%%%%%%%%%%%%%%%%%%%%%%%%%%%%%%%%%%%%%%%%%%%%%%

\paragraph*{Field theory.}
The properties of the quantum critical regime at finite temperature can be described in terms of an imaginary-time Bose-Fermi action,
%is characterized by two low-energy degrees of freedom: The critical fluctuations of the AFM order-parameter $\boldsymbol \Phi_{\ve i}$ and
%the composite fermions $\psi_\sigma$, which dominate the density of single-fermion excitations at $\Jkc$~\cite{liu2022}. From a theory perspective, this can be described in terms of a Bose-Fermi action in the imaginary times framework
\begin{align}
\begin{split}
& S   = \frac{1}{2} \sum_{\Omega_n,\ve q_2} \boldsymbol \Phi_{\Omega_n,\ve q_2} \cdot D_0^{-1}(\Omega_n,\ve q_2) \boldsymbol \Phi_{-\Omega_n,-\ve q_2} \\
& + \!\!\!\!\!\! \sum_{{\omega_n,\ve k_2, \sigma}} \!\!\!\!\!  \bar{\psi}_{\omega_n,\ve k_2,\sigma} [G_{\psi\psi}^{(0)}]^{-1} \psi_{\omega_n,\ve k_2,\sigma}\! +\! 2\geff\!\! \int_0^\beta\!\!\!\! d\tau\!\! \sum_{\ve i} \ve S^\psi_{\ve i}(\tau)\! \cdot \! \boldsymbol \Phi_{\ve i}(\tau).
% + \frac{\geff}{2}\int_0^\beta \sum_{\ve i,\sigma,\sigma'} \bar{\psi}_{\ve i,\sigma}(\tau) \boldsymbol \sigma_{\sigma \sigma'} \psi_{\ve i,\sigma'}(\tau) \cdot \boldsymbol \Phi_{\ve i}(\tau) \, .
 \end{split}
\end{align}
The critical fluctuations of the AFM order parameter are incorporated by the real bosonic field $\boldsymbol \Phi$ 
%is a real bosonic field 
with Matsubara frequencies $\Omega_n$. The Grassmann fields $\psi, \bar \psi$ represent the composite fermions, with Matsubara frequencies $\omega_n$, which dominate the fermion density of states at criticality~\cite{liu2022}. 
The effective coupling $\geff$ considers only the most relevant interaction between the AFM fluctuations and the spin density of the composite fermions $\ve S^\psi_{\ve i}=1/2\sum_{\sigma,\sigma'}\bar{\psi}_{\ve i,\sigma} \boldsymbol \sigma_{\sigma \sigma'} \psi_{\ve i,\sigma'}$.
%Including only their most relevant interaction
%the critical properties due to interactions between the gapless  and the spin density of the composite Fermions. These can be described in terms of an effective interaction potential
%$H_{\text{int}} = g_{\text{eff}} \sum_{\mathbf{R}_2} \psi^\dagger_{\ve R_2,\sigma} %\boldsymbol{\sigma}_{\sigma \sigma'} \psi_{\ve R_2, \sigma'} \cdot \ve \Phi_{\ve R_2} \, .$
%providing the basis for a field theoretic analysis.
Similar models have been discussed in various contexts, describing, e.g., conduction electrons coupled to ferromagnetic~\cite{chubukov2004InstFM, chubukov2009ConsFL}, Ising-nematic~\cite{metzner2003soft,dellanna2006, metlitski2010quantum1}, or antiferromagnetic~\cite{abanov2003SpinFermion, metlitski2010quantum2, schlief2017exact} order parameters, as well as to a U(1) gauge field~\cite{palee1989U1, altshuler1994, sslee2009U1}. In two dimensions, these setups can potentially host non-Fermi liquid states, characterized by the lack of well-defined quasiparticles~\cite{2007_HvL_RMP,lee2018recent}.
%While being of interest in their own right these unconventional 
%
Importantly, the critical fluctuations may alter the behavior of the fermion excitations beyond the well-established Hertz-Millis theory~\cite{hertz1976,millis1993,moriya2012spin}, which considers only the dressing of the order-parameter fluctuations by correlations of essentially noninteracting electrons.
\begin{figure}[t]
\begin{center}
\includegraphics[width=0.9\columnwidth]{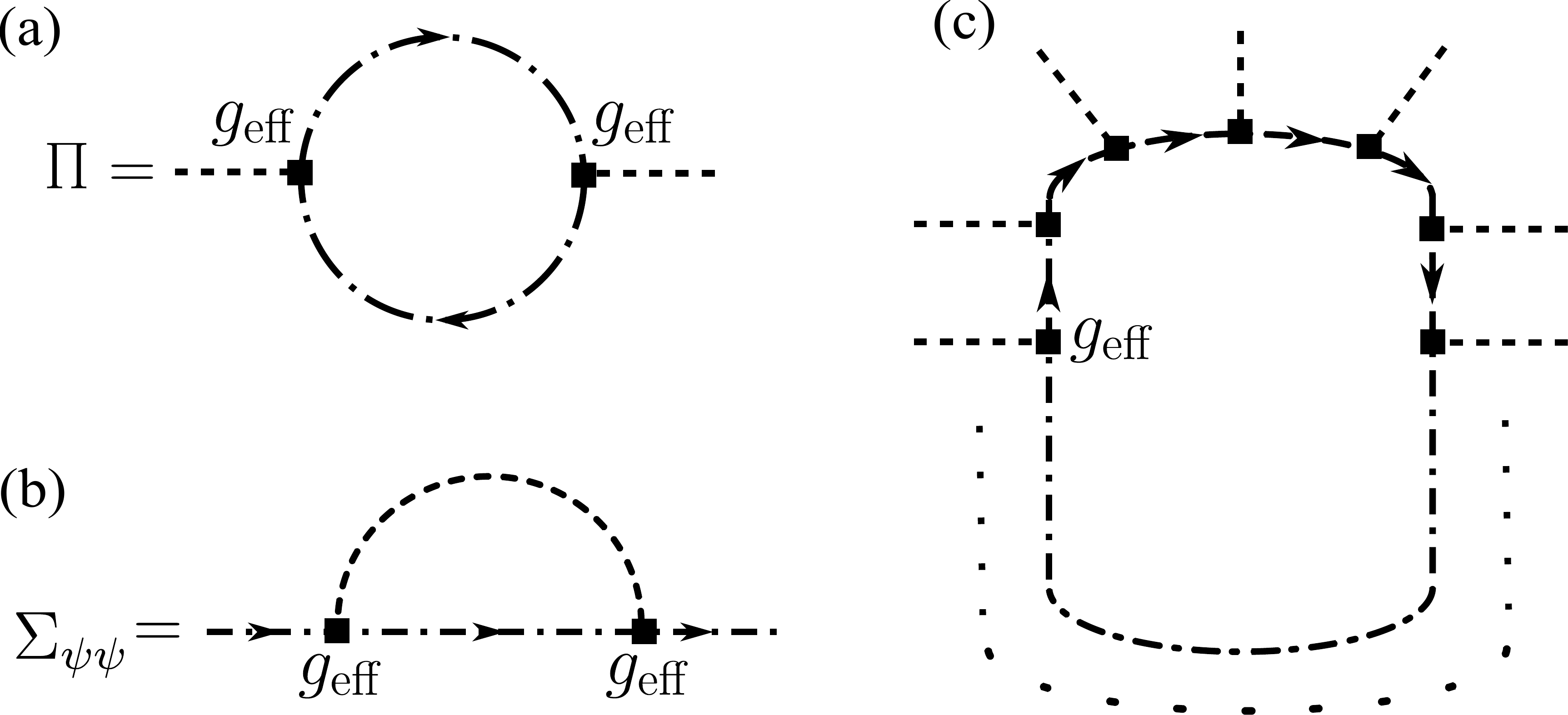}
\caption{Self-energy for (a) order-parameter and (b) fermion fields. Dashed propagators correspond to $D$ and dotted-dashed one to $G_{\psi\psi}$. Squares denote the effective interaction strength $g_\text{eff}$. (c) Higher-order order-parameter vertices.}
\label{fig:FeynmanDiag}
\end{center}
\end{figure}
In the above equation, the bare fermionic propagator includes the mean-field hybrization effects,
\begin{align}\label{eq:defTmf}
G^{(0)}_{\psi \psi}(\omega_n,\ve k_2)\! = \!\frac{4}{\Jk^2} \tilde T_\mf(\omega_n,\ve k_2)\! =\! \frac{4 \Jk^{-2}}{v_\mf^{-2} i \omega_n - \tilde\epsilon_{\ve k_2}-g_0(\omega_n,\ve k_2)}
%\frac{4 \Jk^{-2}}{\dfrac{i \omega_n - \tilde\epsilon_{\ve k_2}}{\vmf^2}-g_0(\omega_n,\ve k_2)} \, ,
\end{align}
in terms of the mean-field transition matrix $ \tilde T_\mf$, encoding the out-of-plane scattering of conduction electrons off the spin layer.
Here, we introduce $g_{0}(\omega_n,\ve k_2) = 
%L^{-1} \sum_{k_z}(i\omega_n -\epsilon_{\ve k})^{-1} =
1/(\sqrt{i \omega-\epsilon_{\ve k_2}+2t} \sqrt{i \omega-\epsilon_{\ve k_2}-2t})$ 
with in-plane dispersion $\epsilon_{\ve k_2} = -2t(\cos k_x + \cos k_y)$
while $v^{-2}_\mf i \omega - \tilde\epsilon_{\ve k_2}$ incorporates the Kondo resonances from screening the local moments with weights $\vmf^2$ and dispersion $\tilde\epsilon_{\ve k_2} =  (t'/t) \epsilon_{\ve k_2} $.
The derivation of Eq.~\eqref{eq:defTmf}
% along the lines of Ref.~\cite{borda2007} for Kondo impurities 
and a comparison with QMC data are given in the Supplemental Material (SM)~\cite{SM}.
Furthermore, we model the bare antiferromagnetic susceptibility with the standard asymptotic form
$D^{(0)}(\Omega_n,\ve q_2) = (\Omega_n^2 + c_B^2(\ve q_2-\ve Q)^2 + M_0^2)^{-1}$,
with instability wavevector $\ve Q = (\pi,\pi)$ and boson velocity $c_B$. $M_0^2$ describes the bare spectral gap.
Interactions dress the propagators via self-energies $\Pi$ and $\Sigma_{\psi\psi}$ as 
$D(\Omega_n,\ve q_2) = (D^{(0)}(\Omega,\ve q_2)^{-1}+\Pi(\Omega_n,\ve q_2))^{-1}$ and
$G_{\psi\psi} = 4/\Jk^2(\tilde T_\mf^{-1}-(4/\Jk)^2\Sigma_{\psi\psi})^{-1}$.
At the QCP, the gap vanishes, $M^2 \equiv M^2_0+\Pi(\Omega=0,\ve Q)=0$, while a finite value $M^2>0$ ($M^2 < 0$) drives the system into the paramagnetic (antiferromagnetic) heavy-fermion metal.
In the following, we perform a one-loop analysis that captures the most important interaction effects, see Fig.~\ref{fig:FeynmanDiag}(a,b) for corresponding diagrams.
%
% and translate to the one-loop expressions for the bosonic:
%\begin{align}\label{eq:oneloopPi}
%\Pi(\Omega_n,\ve q_2) = \frac{16 \geff^2}{ \Jk^{4}\beta}\!\!\! \sum_{\omega_m,\ve k_2}\!\!\! \tilde %T_\mf(\omega_m,\ve k_2) \tilde T_\mf(\omega_m+\Omega_n,\ve k_2+\ve q_2)
%\end{align}
%and fermionic self-energy, where we insert the dressed $D$
%\begin{align}\label{eq:oneloopPsi}
%\begin{split}
%\Sigma_{\psi\psi}(\omega_n,\ve k_2) = \frac{4\geff^2}{\Jk^2\beta} \sum_{\Omega_m,\ve q_2}\!\!\!%D(\Omega_m,\ve q_2) T_\mf(\omega_n+\Omega_m,\ve k_2+\ve q_2).
%\end{split}
%\end{align}

\paragraph*{Results at $\mathit{T=0}$.} 
We start with the behavior at the QCP.
%Let us start with the analysis of the correlations in the ground state where  imaginary frequencies are continuous variables. 
Note that $G^{(0)}_{\psi\psi}$ does not exhibit the typical free-particle poles  
%$(i \omega -\xi_{\ve k})^{-1}$ at small energies 
in the vicinity of the Fermi surface. Instead, only a discontinuity appears at in-plane momenta $\ve k_2$ within the projected 2D Fermi surface, shown in Fig.~\ref{fig:QMCmatch}(b). 
%which results from $g_{0}(\omega \to 0,\ve k_2) \to -i\text{sgn}(\omega)(4t^2-\epsilon_{\ve k_2}^2)^{-1/2}\theta(2t-|\epsilon_{\ve k_2}|)$. 
This more regular infrared behavior has important consequences for the critical properties.  
%is finite at small Matsubara frequencies for all $\ve k_2$, even in the ground state where the limit of the continuous variable $\omega\to 0$ can be taken. The only exceptions are the van Hove singularities of the effective one-dimensional band edge in $g_0$ where $T_\mf$ vanishes. This is in marked contrast --with important consequences for the quantum critical behavior-- to the homogeneous system discussed above where the single-electron Green's functions exhibit infrared singularities on the FS.
%The evaluation of Fig.~\ref{fig:FeynmanDiag}(a) yields the asymptotic form~\cite{SM}
%$\Pi(\Omega \to 0,\ve q_2 \to \ve Q) = \Pi(0,\ve Q) + \alpha |\Omega|$, 
%$\Pi(\Omega,\ve Q) = \Pi(0,\ve Q) + \alpha |\Omega| + \mathcal O(|\Omega^2|)$, 
%with the nonanalyticity parametrized by the nonuniversal prefactor $\alpha$ representing Landau damping, in a form typically encountered in AFM metallic QCPs~\cite{hertz1976}. 
%Here, the prefactor $\alpha$ is nonuniversal and contains an integral over the entire 2D Brillouin zone of in-plane momenta. 
%
The evaluation of Fig.~\ref{fig:FeynmanDiag}(a) leads to the dressed boson propagator at criticality as~\cite{SM}
%
%Considering the effects of $g_{\text{eff}}$ on a perturbative level gives rise to the self-energies depicted in Fig.~\ref{fig:FeynmanDiag}.
%The dynamic susceptibility is dressed via $D(\Omega,\ve q_2) = (D^{(0)}(\Omega,\ve q_2)^{-1}-\Pi(\Omega,\ve q_2))^{-1}$ with the additional criterion $M^2_0-\Pi(\Omega=0,\ve Q)=0$ to establish the QCP with a gapless critical susceptibility.
%Evaluating the one-loop diagram $\Pi(\Omega,\ve q_2) = -16 g^2_{\text{eff}} J_K^{-4} \cdot \int d\omega/(2\pi) \int_{2d BZ} d^2k_2/(2\pi)^2 T_\mf(\omega,\ve k_2,R_z=0)T_\mf(\omega+\Omega,\ve k_2+\ve Q+\ve q_2,R_z=0) $ provides the Landau damped magnetic susceptibility at the QCP (see SM for details)
%
\begin{align}\label{eq:Ddressed}
D(\Omega,\ve q_2) = \frac{1}{c_B^2 (\ve q_2-\ve Q)^2 + \alpha |\Omega|}  \, ,
\end{align}
with the nonanalyticity, parametrized by the nonuniversal prefactor $\alpha$, representing Landau damping, in a form typically encountered in AFM metallic QCPs~\cite{hertz1976}. 
It implies a dynamical exponent $z=2$, in agreement with the QMC results~\cite{liu2022}.
%
%Turning to the fermionic sector interaction effects beyond the MF-level occur via the self-energy $\delta \Sigma_{\psi \psi}$ that gives rise to the dressed Green's function $G_{\psi\psi} = ((\frac{4}{J_K^2}T_\mf)^{-1}-\delta\Sigma_{\psi\psi})^{-1} = \frac{4}{J_K^2}(T_\mf^{-1}-\frac{4}{J_K}^2\delta\Sigma_{\psi\psi})^{-1}$.
For the fermion self-energy, we find from Fig.~\ref{fig:FeynmanDiag}(b), using the dressed boson propagator~\cite{SM}
\begin{align}\label{eq:resImSigmaT0}
\begin{split}
&\Sigma_{\psi\psi}(\omega,\ve k_2) \!=\! - i  \gamma(\epsilon_{\ve k_2}) \omega \log\left(\frac{e ^2 \Lambda^2}{\alpha|\omega|}\right) ,
\end{split}
\end{align}
%
%because of the regular low-frequency behavior of $T_\mf$. 
which holds for $\ve k_2$ within the projected 2D Fermi surface.
%
% [Fig.~\ref{fig:QMCmatch}(b)]. 
%
Here, $\gamma(\epsilon_{\ve k_2})$ denotes a nonuniversal momentum-dependent prefactor and 
$\Lambda$ is the momentum cutoff.
%
%\lj{Does $\gamma(\ve k_2)$ depend on both components of $\ve k_2$ independently, or only on $\epsilon_{\ve k_2}$? I.e., $\gamma(\ve k_2) \to \gamma(\epsilon_{\ve k_2})$?}
%
% beyond which the expanded momentum dependence in $D(\Omega,\ve q_2)$ around the K point breaks down. 
Analytic continuation $\Sigma_{\psi \psi} (i \omega \to \omega + i0^+)$ to real frequencies yields the retarded self-energy $\Sigma_{\psi\psi}^R$ whose imaginary part encodes the decay rate of the single-particle excitations. For the Kondo heterostructure we find $-\Im \Sigma_{\psi \psi}^R(\omega,\ve k_2) \sim |\omega|$, which implies marginal-Fermi-liquid scaling~\cite{varma1989}.
% in the sense that the fluctuation-induced decay rate scales linearly with excitation energy. 
The unusual scaling originates from the discontinuity of $G^{(0)}_{\psi\psi}$ and differs from the decay rate $\sim |\omega|^{1/2}$ obtained in case of a critical AFM order parameter coupled to a 2D Fermi surface~\cite{abanov2003SpinFermion}.
Below, we will see how the marginal-Fermi-liquid behavior is reflected in the temperature dependence of the resistance of the heterostructure. 

%Next, we address the universality class of the QCP in the Kondo heterostructure.
In addition to these perturbative results, we can make precise statements about the universality class of bosonic sector of the Kondo heterostructure, which typically is a hard problem in metallic quantum criticality. As will be shown below, the resulting scaling relation influence the marginal Fermi liquid at finite temperatures.
The problem of an AFM QCP coupled to a 2D Fermi surface with poles in the electron propagators $(i\omega-\xi_{\ve k_2})^{-1}$ was studied by Abanov and Chubukov~\cite{abanov2004anomalous}. They showed that the class of diagrams in Fig.~\ref{fig:FeynmanDiag}(c), consisting of a single fermion loop with $2n\geq 4$ external order parameter fields $\ve{\Phi}$ attached, acquire universal infrared singularities. These make all the associated higher order vertex functions $\Gamma^{(2n)}$ marginally relevant under the renormalization group flow, which in total leads to a nonzero anomalous dimension of the order parameter field even at the effective upper critical dimension $d+z=2+2=4$. The bosonic part of the criticality of the Kondo heterostructure is characterized by a 2D AFM order parameter, too, but the fermion propagator $G_{\psi\psi}$ here is more regular at small energies. Consequently, the infrared behavior of the one-loop diagrams is also regularized and the vertex functions $\Gamma^{(2n)}$ become RG irrelevant instead of marginally relevant. To study the critical properties, it suffices, therefore, to consider the Hertz-Millis action~\cite{hertz1976,millis1993} for the coarse grained order parameter $\boldsymbol \Phi(\tau,\ve x_2)$ in the spin layer
%In addition, the fact that $T_\mf(\omega \to 0,\ve k_2)$ is well-defined
%allows to determine the universality class of the QCP in the Kondo heterostructure: The most relevant higher-order bosonic vertex diagrams $\Gamma^{(2n)}$ (see Fig.~\ref{fig:FeynmanDiag}), consist of a single $\psi_\sigma$ loop and $2n\geq 4$ external order parameter fields $\ve \Phi(\Omega_l,\ve q_{2,l}), l=1,...,2n$. Since the internal line does not show any singular behavior in the infrared the vertex functions approach nonuniversal but finite values $\Gamma^{(2n)} \to u^{(2n)}$, even for the choice $\Omega_l=0$ and $\ve q_{2,l}=\ve Q$ for all $l$. As a result, all $\Gamma^{(2n)} \mathbf{\Phi}^{2n}, 2n\geq 6$ are irrelevant and the critical behavior is fully captured by the standard effective action
\begin{align}
S[\ve \Phi] =\frac{1}{2}\!\int_{\Omega,\ve q_2}\!\!\!\!\boldsymbol{\Phi}_{\Omega,\ve q_2}D^{-1}(\Omega,\ve q_2) \boldsymbol{\Phi}_{-\Omega,-\ve q_2} + u^{(4)}\!\! \int_{\ve x_2 \tau}\!\! \!\!\boldsymbol \Phi^4(\tau,\ve x_2)\, ,
\end{align}
with $D(\Omega,\ve q_2)$ from Eq.~\eqref{eq:Ddressed} and only quartic interactions. As a result, the QCP of the Kondo heterostructure belongs to the Hertz-Millis universality class in $d=2,z=2$ with well-known scaling relations~\cite{millis1993}. In particular, these affect the behavior of the marginal-Fermi-liquid self-energy $\Sigma_{\psi\psi}$ at finite temperatures, which will be discussed next.
\begin{figure*}[t]
\includegraphics[width=\textwidth]{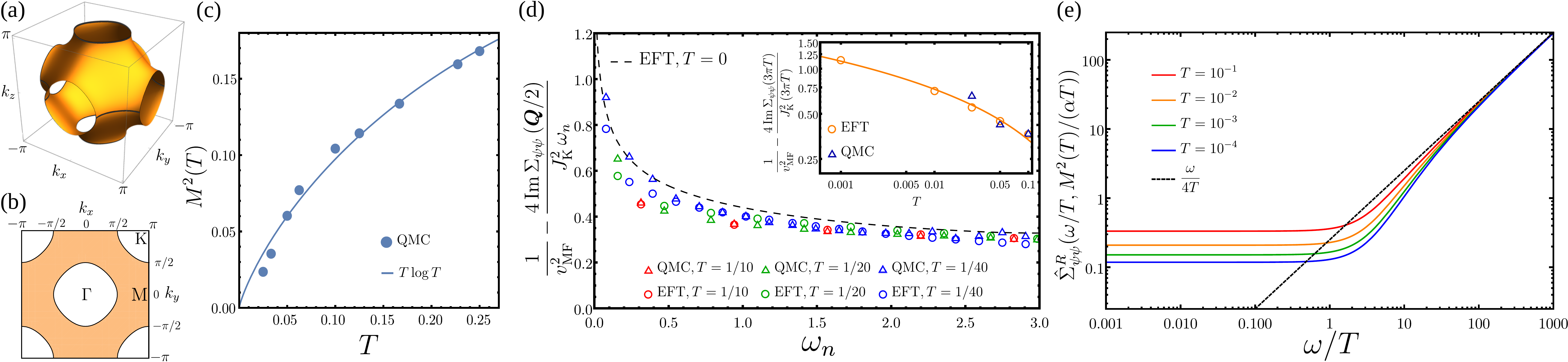}
\caption{
(a)  3D Fermi surface. 
(b) Projected 2D Fermi surface.
(c) Thermal gap as function of temperature from Hertz-Millis asymptotics $M^2(T)\sim T \log T$ (solid line) and QMC at linear system size $L=12$ (dots). 
%<<<<<<< HEAD
%(d) Imaginary part of fermionic self-energy as function of Matsubara frequency from effective field theory (triangles) and $L=12$ QMC (dots) for different temperatures. The inset shows the variation at the Matsubara frequency $\omega_n = 3\pi T$ as function of temperature, which follows a logarithmic asymptotics (solid line).
%=======
(d) Imaginary part of fermionic self-energy as function of Matsubara frequency from effective field theory (triangles) and $L=12$ QMC (dots) for different temperatures. The black dashed line corresponds to the result at $T=0$. The inset shows the variation at the Matsubara frequency $\omega_n = 3\pi T$ as function of temperature, which follows a logarithmic asymptotics (solid line).
%\lj{Replace label ``Hertz-Millis'' by ``$T \log T$'' or similar}
%>>>>>>> 1de1670474e1c4c3844ef56a200d7491e0a1e241
%\lj{Add lines between triangles as guides to the eyes?}
%\lj{Simplify label $\tilde\Sigma_{cc}(\ve k_2 = (\pi/2,\pi/2)) \to \Sigma_{\psi\psi}(\ve Q)$?}
%\lj{Increase font size of labels}
%\lj{Adapt coloring of data points to allow better distinction between different temperatures (red vs brown hardly distinguishable)}
(e) Imaginary part of retarded fermionic self-energy as function of real frequency $\omega$ from effective field theory.
}
\label{fig:QMCmatch}
\end{figure*}

%\begin{figure*}[t]
%\includegraphics[width=0.33\textwidth]{FS+M(T).pdf}\hspace*{5pt}%
%\includegraphics[width=0.3\textwidth]{plotIm157157.pdf}\hspace*{5pt}%
%\includegraphics[width=0.3\textwidth]{realFreqFull.pdf}
%\caption{%
%(a)  3D Fermi surface. 
%
%(b) Projected 2D Fermi surface.
%
%(c) Thermal gap as function of temperature from Hertz-Millis asymptotics $M^2(T)\sim T \log T$ (solid line) and QMC at linear system size $L=12$ (dots). 
%
%(d) Imaginary part of fermionic self-energy as function of Matsubara frequency from effective field theory (triangles) and $L=12$ QMC (dots) for different temperatures.
%
%\lj{Add lines between triangles as guides to the eyes?}
%\lj{Simplify label $\tilde\Sigma_{cc}(\ve k_2 = (\pi/2,\pi/2)) \to \Sigma_{\psi\psi}(\ve Q)$?}
%\lj{Increase font size of labels}
%\lj{Adapt coloring of data points to allow better distinction between different temperatures (red vs brown hardly distinguishable)}
%\lj{Question: Why do we plot $\frac{1}{\omega}\Im\Sigma^{-1}$ instead of, say, just $\Im\Sigma$?}
%
%(e) Imaginary part of retarded fermionic self-energy as function of real frequency $\omega$ from effective field theory.
%
%\lj{Adapt color scheme to agree with everyday scheme (red = high $T$, blue = low $T$)?}
%
%\bof{add T=0 and extrapolation in a good way; improve overall appearance}
%
%}
%\label{fig:QMCmatch}
%\end{figure*}

\paragraph*{Finite temperature and comparison with QMC.}
Next, we discuss our finite-temperature results and compare with the QMC data in the quantum critical regime. 
We follow the procedure developed in Ref.~\cite{klein2020} and compute $\Sigma_{\psi\psi}(\omega_n,\ve k_2)$ self-consistently at the one-loop level, to capture thermal effects properly, while we use numerical input from the QMC simulations to set the values of the nonuniversal parameters characterizing the dressed propagators.
For instance, in case of the boson propagator, in the vicinity of the instability wavevector $\ve Q = (\pi, \pi)$, we expect an asymptotic behavior
\begin{align}\label{eq:DfiniteT}
D(\Omega_n,\ve q_2) = \frac{1}{d_0^{-1} \Omega_n^2 + c_B^2 (\ve q_2-\ve Q)^2 + M^2(T)+\alpha |\Omega_n|} \, ,
\end{align}
%
%in the vicinity of the K point 
%
where we have introduced, in addition to the Landau damping, an analytic frequency dependence with nonuniversal prefactor $d_0^{-1}$, and an effective finite-temperature gap $M^2(T)$, realizing the thermal cutoff, with $M^2(0)=0$. In general, $M^2(T)$ may realize a rather rich behavior as function of temperature~\cite{frank2020}. Here, however, Hertz-Millis scaling dictates  the form $M^2(T) \sim  T \log T$, which agrees well with the QMC results, see Fig.~\ref{fig:QMCmatch}(c). This allows us to fix all nonuniversal parameters in the bosonic sector uniquely.
%
%For details how the numerical prefactors and the remaining parameters $\alpha$, $c_B$ and $d_0$ are obtained, see Ref.~\cite{SM}. 
Similarly, the renormalized hopping amplitude $t'/t$ of the Kondo resonances can be extracted directly from the QMC results for the propagator of the conduction electrons. The weight of the resonances $v^2_\mf$, as well as the fermion-boson coupling $\geff$, can be obtained by matching the self-consistent result for $\Sigma_{\psi\psi}(\omega_n, \ve Q)$ at a fixed temperature, which we choose as $T/t=1/10$, see SM \cite{SM} for details and, in particular, Table~\ref{numVal} therein for a summary of all parameters. To compare the effective field theory we use the identity $\Im [\tilde T(\omega_n,\ve k_2) -g_0(\omega_n,\ve k_2)] = v_\mf^{-2} \omega_n - 4 \Jk^{-2} \Im \Sigma_{\psi\psi}(\omega_n,\ve k_2)$. The left-hand side can be extracted directly from the QMC data while the right-hand side is obtained from the effective field theory. To enhance the resolution we divide by $\omega_n$.
%
%To evaluate $\tilde T_\mf$ from Eq.~\eqref{eq:defTmf} we use a similar procedure: $\tilde{\epsilon}_{\ve k_2}=t'/t \epsilon_{\ve k_2}$ is determined from the QMC data but it turns out that the self-energy $\Im \Sigma_{\psi \psi}$ gives pronounced corrections to $\Im \tilde T_\mf^{-1} \sim i \omega_n/v^2_\mf$.
%Since there is no controlled way to separate the mean-field contribution from other interaction effects we are forced to keep $\vmf$ as free parameter. 
%Together with the effective coupling $\geff$, our one-loop theory has thus only two free parameters.   
%To compare the behavior of the composite fermions at finite temperature with the numerics, we calculate the self-energy $\Sigma_{\psi\psi}$ from Eq.~\eqref{eq:oneloopPsi} self-consistently to capture thermal effects properly according to Ref.~\cite{klein2020}. This means we use $D$ with the prefactors from the fitting procedure and the dressed T-matrix
%We set their values by matching the converged result for $\Sigma_{\psi \psi}\left(\omega_n,\ve k_2= (\pi/2,\pi/2)\right)$ to the QMC simulations at the largest temperature. After the elimination of all free parameters, we compare the results for $\Sigma_{\psi\psi}$ to the QMC data at lower temperatures in Fig.~\ref{fig:QMCmatch}(b) and to other in-plane momenta~\cite{SM}.
%
The resulting agreement between the effective field theory and the QMC data for the fermionic self-energy as function of Matsubara frequency $\omega_n$ for different temperatures and linear system size $L=12$ is remarkable, see Fig.~\ref{fig:QMCmatch}(d).
This includes the correction to the linear behavior in general, as well as the pronounced increase at the smallest $\omega_n$.
Importantly, the effective field theory allows us to extrapolate to lower temperatures down to $T=0$. As revealed in Fig.~\ref{fig:QMCmatch}(d) this limit is approached only logarithmically with decreasing temperature. 
%\bof{Make better statement} Simultaneously, the ground state exhibits a logarithmic frequency dependence, as expected from Eq.~\eqref{eq:resImSigmaT0}. 
%
%Further details on the computations can be found in Ref.~\cite{SM}.

%%%%%%%%%%%%%%%%%%%%%%%%%%%%%%%%%%%%%%%%%%%%%%%%%%%%%%%%%%%%%%%%%%%%%%%

\paragraph*{Transport.} 
Finally, we address the question how the marginal Fermi liquid can be observed experimentally. Since the Kondo interactions are restricted to the spin layer, they do not affect the
bulk thermodynamic properties of the conduction electrons. Transport across the spin plane, however, is sensitive to the presence of the local moments. In the following, we consider a simple setup in which the region above the spin plane, $R_z>0$, is subject to a static homogeneous electric potential $e V>0$. By symmetry, only a finite stationary current density along the positive out-of-plane direction
$\<\jmath^z\>$ will be generated.  
%As a result, the current density operator in the out-of-plane direction will acquire 
%a positive expectation value $\<j^z\>$, which is independent of $R_z$ in a stationary state, whereas, by symmetry, no current will flow parallel to the plane. 
Within linear response, the conductivity $\sigma$ is given by Ohm's law $\<\jmath^z\> = \sigma V/(2a)$, where $a=1$ is the lattice constant~\footnote{This setup neglects intrinsic defects within the metals, which break translation invariance on the microscopic level and affect the homogeneity of $\<\jmath^z  \>$. For good metals these effects are expected to be subleading for large enough sample sizes. Furthermore, intrinsic interactions of the electrons within the metallic regions $R_z \neq 0$  are not considered. These lead to an increase of $\rho(T)$ with increasing temperature as depicted schematically in Fig.~\ref{fig:SysTrans}(b)}.
The calculation of $\sigma$ is tremendously simplified by the conservation of in-plane momenta, which allows us to theoretically decompose the Kondo heterostructure into $L^2$ independent one-dimensional scattering problems, labeled by $\ve k_2$. Each of these realizes an interacting quantum dot in the layer $R_z=0$, connected to two identical leads formed by the noninteracting regions $R_z<0$ and $R_z>0$, respectively. For a fixed $\ve k_2$, the conductivity can therefore be obtained via the Meir-Wingreen formalism~\cite{meir1992}.
% for interacting mesoscopic systems with two noninteracting leads. Application of their method and summing over $\ve k_2$ yields the total conductivity in units  (see~\cite{SM} for details):
Summing over $\ve k_2$ yields the total conductivity
\begin{align}\label{eq:defSigma}
\sigma = -4 \pi  e^2 t^2 \int \frac{d^2 k_2}{(2\pi)^2} d\omega  a_0(\omega, \ve k_2)
\left. n_F'(\omega)\right|_{\mu=0} a(\omega,\ve k_2) \, .
\end{align}
In the above, $n_F'(\omega)=dn_F(\omega)/d\omega$ denotes the derivative of the Fermi distribution and $a_0(\omega, \ve k_2) = -\pi^{-1} \Im g_0(i \omega_n \to \omega+ i 0^+,\ve k_2)$ is the free one-dimensional density of states of the leads. The only genuine interaction contribution is given by the dressed local spectral function of the conduction electrons $a(\omega,\ve k_2) = -\pi^{-1} \Im g(i\omega_n \to \omega + i 0^+,\ve k_2)$ that derives from the dressed local Green's function Eq.~\eqref{eq:defgfull}. 
At $T=0$, the conductivity attains a nonuniversal value $\sigma(T=0)$, which originates from scattering off conduction electrons from the Kondo-screened local moments. 
However, the deviation $\delta \sigma(T) = \sigma(T) -\sigma(0)$ shows the same temperature dependence as the imaginary part of the analytically-continued retarded self-energy $\Im\Sigma^R_{\psi\psi}$, implying that the conductivity is sensitive to the nature of the composite fermion excitations and their marginal-Fermi-liquid behavior. The diagrammatic evaluation~\cite{SM} reveals that the frequency and temperature dependence of the retarded self-energy can expressed as  $\Im\Sigma^R_{\psi\psi}(\omega,\ve k_2) \sim T \hat{\Sigma}_{\psi\psi}^R(\omega/T,M^2(T)/\alpha T)$, where $\hat\Sigma^R_{\psi\psi}$ is a momentum-independent, universal scaling function, see Fig.~\ref{fig:QMCmatch}(e). 
In the quantum critical regime, we obtain two different limiting behaviors: (i) For large frequencies $|\omega|/T \gg 1$, we have 
%$\bar{\Sigma}_{cc}^R(|\omega|/T,M^2(T)/\alpha T) \sim |\omega|/T$, 
%
$\hat{\Sigma}_{\psi\psi}^R \sim |\omega|/T$,
such that $\Im \Sigma^R_{\psi\psi} \sim |\omega|$ recovers the marginal-Fermi-liquid behavior from the QCP, as expected. 
(ii) For small frequencies $|\omega|/T \ll 1$, we obtain a constant
%$\bar{\Sigma}_{cc}^R(|\omega|/T \ll 1,M^2(T)/\alpha T,\ve k_2) \sim T^2/M^2(T) \sim T/\log T$ 
$\hat{\Sigma}_{\psi\psi}^R(0,M^2(T)/\alpha T)$, which scales like $T^2/M^2(T) \sim T/\log T$ in the limit of small temperatures.
%as a consequence of Hertz-Millis scaling. 
%
Since the overall scale of the self-energy is set by the temperature, we obtain $\delta \sigma(T) \sim T$ up to logarithmic corrections, see Fig.~\ref{fig:sigma}. Converted to the resistivity $\rho(T)=1/\sigma(T)$, these results imply a linear decrease with temperature for temperatures below the coherence scale $T_\text{coh}$, as schematically indicated in Fig.~\ref{fig:SysTrans}(b). 
In the paramagnetic heavy-fermion metal at sufficiently large $\Jk$, we find instead the Fermi-liquid self-energy $\Im \Sigma^R(\omega,\ve k_2) \sim \max(\omega^2,T^2)$, resulting in $\delta\sigma(T) \sim T^2$, implying the conventional form of $\rho(T)$ at low temperatures~\cite{nozieres74}, as also indicated in Fig.~\ref{fig:SysTrans}(b). 

%%%%%%%%%%%%%%%%%%%%%%%%%%%%%%%%%%%%%%%%%%%%%%%%%%%%%%%%%%%%%%%%%%%%%%%
\begin{figure}[b]
%\begin{center}
\includegraphics[width=\columnwidth]{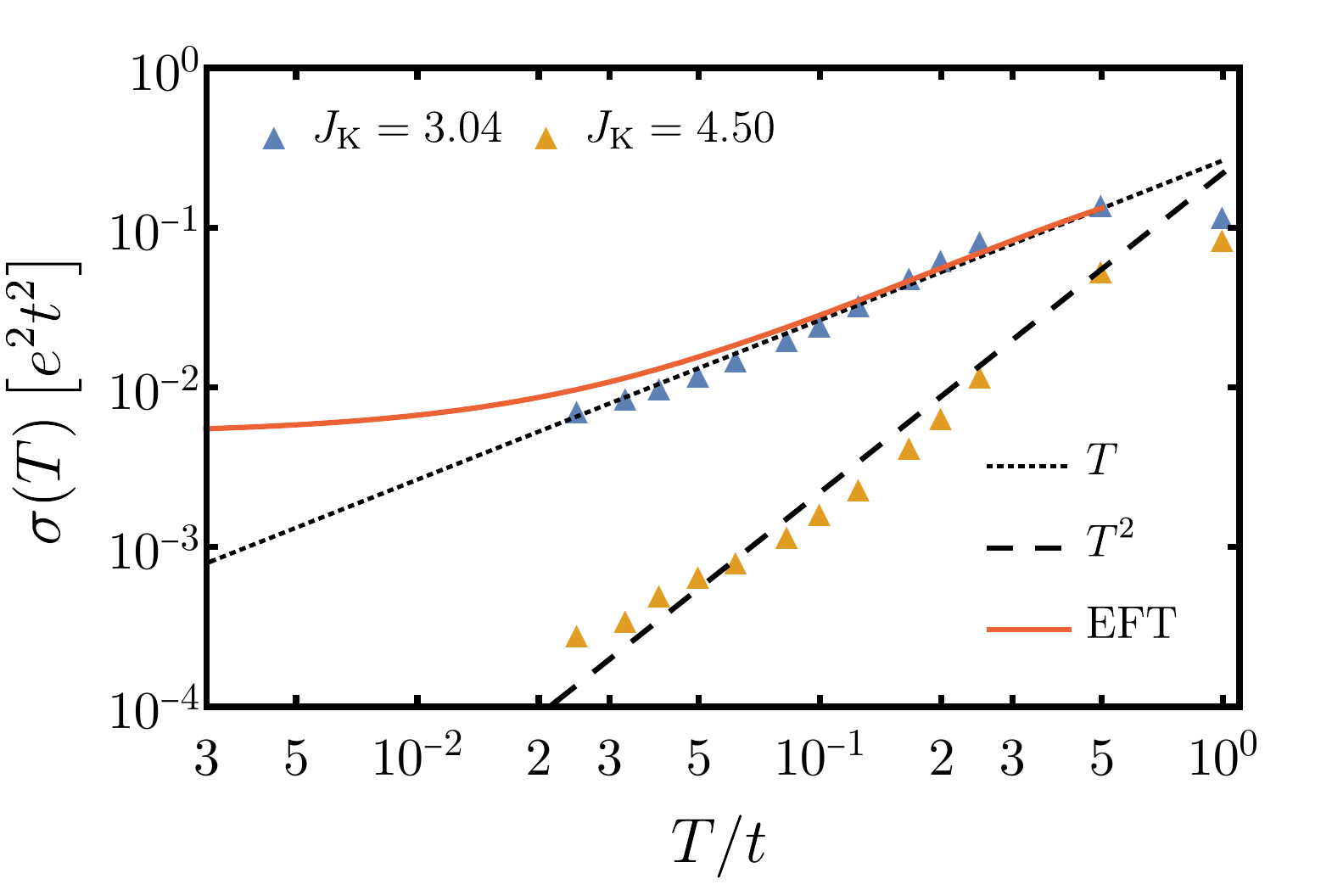}
\caption{%
Conductivity as function of temperature from effective field theory at criticality (red solid line) and QMC simulations for $\Jk = 3.04 t \approx \Jkc$ (blue triangles) and $\Jk = 4.50 t > \Jkc$ (yellow triangles) for linear system size $L=10$. 
Black dotted (dashed) line indicates the linear (quadratic) marginal-Fermi-liquid (Fermi-liquid) scaling.
%
%\bof{Say something about AFM phase?}
%\lj{I suggest to move the data points for $J<\Jk$ to the SM.}
%
%\bof{It is hard to distinguish dashes and dots.}
%\lj{Agreed -- can this be improved?}
%
%\lj{Extend plot to smaller values of $T$?}
}
\label{fig:sigma}
%\end{center}
\end{figure}
%%%%%%%%%%%%%%%%%%%%%%%%%%%%%%%%%%%%%%%%%%%%%%%%%%%%%%%%%%%%%%%%%%%%%%%

The conductivity may be also accessed from the QMC simulations without numerical analytic continuation, provided that one approximates $n'_F(\omega) \approx -\delta(\omega)$.
Such replacement neither affects the temperature scaling nor introduces a large numerical error, see \cite{SM}. 
The value of $a(\omega=0,\ve k_2)$ at a single real frequency can then be obtained from the QMC data at imaginary times in a controlled way for $T \to 0$.
%
%\bof{Maybe we should add a few lines to the SM? or better shift some information to the SM}
%
Figure~\ref{fig:sigma} shows that the QMC simulations indeed display the expected linear (quadratic) scaling for $\Jk \approx \Jkc$ ($\Jk > \Jkc$). Moreover, at criticality, we even find very good quantitative agreement between the QMC data and the effective field theory. This is also reflected in the resistance as function of temperature at the quantum critical point, as presented in Fig.~\ref{fig:SysTrans}(c).

\paragraph*{Summary.}
%
%We have identified a quantum critical point in a model for a Kondo heterostructure, which realizes a novel instance of a marginal Fermi liquid, characterized by a linear temperature dependence of the electric resistance measured across the magnetic layer.
%
We have identified a quantum critical point in a model for a Kondo heterostructure, which realizes a novel instance of a marginal Fermi liquid, characterized by a linear temperature dependence of the electrical resistivity measured across the magnetic layer.
%
%Our results call for careful low-temperature transport experiments on appropriate heterostructures made of a single magnetic layers embedded in bulk metals, that can be tuned near a quantum critical point. 
%
Our results call for careful low-temperature transport experiments on appropriate heterostructures made of single magnetic layers embedded in bulk metals, that can be tuned close to a quantum critical point.
In this respect, we note that various heavy-fermion compounds, such as bulk CeIn$_3$, feature a pressure-driven quantum phase transition \cite{mathur98}, suggesting that pressurized CeIn$_3$/LaIn$_3$ \cite{shishido10} or related heterostructures could realize some of the physics discussed in this work.
%
%In this respect, we note that various heavy-fermion compounds, such as bulk CeIn$_3$, can be tuned through a quantum phase transition by applying pressure \cite{mathur98}, suggesting that CeIn$_3$/LaIn$_3$ \cite{shishido10} or related heterostructures under pressure could realize some of the physics discussed in this work.
%
%\lj{Any other suggestions for outlook and/or comments, references, ...?}
On a more general note, our work provides a new path to engineer, and characterize in transport experiments, exotic quantum phases of matter using the concept of dimensional mismatch.

%%%%%%%%%%%%%%%%%%%%%%%%%%%%%%%%%%%%%%%%%%%%%%%%%%%%%%%%%%%%%%%%%%%%%%%

%\begin{acknowledgments}

\paragraph*{Acknowledgments.}
FA    and  MV   acknowledge  enlightening conversations  with  T.  Grover   and B.  Danu  on   related  subjects. 
The authors gratefully acknowledge the Gauss Centre for Supercomputing e.V.\ (www.gauss-centre.eu) for funding this project by providing computing time on the GCS Supercomputer SUPERMUC-NG at Leibniz Supercomputing Centre (www.lrz.de),   (project number pn73xu)    
as  well  as  the scientific support and HPC resources provided by the Erlangen National High Performance Computing Center (NHR@FAU) of the Friedrich-Alexander-Universit\"at Erlangen-N\"urnberg (FAU) under the NHR project b133ae. NHR funding is provided by federal and Bavarian state authorities. NHR@FAU hardware is partially funded by the German Research Foundation (DFG) -- 440719683.
This research has been supported by the Deutsche Forschungsgemeinschaft through the W\"urzburg-Dresden Cluster of Excellence on Complexity and Topology in Quantum Matter -- \textit{ct.qmat} (EXC 2147, Project No.\ 390858490), SFB 1170 on Topological and Correlated Electronics at Surfaces and Interfaces (Project No.\  258499086), SFB 1143 on Correlated Magnetism (Project No.\ 247310070), and the Emmy Noether Program (JA2306/4-1, Project No.\ 411750675).
%
%\end{acknowledgments}

\bibliographystyle{shortapsrev4-2}

\bibliography{2d_3d_kondo_prl}

%%%%%%%%%%%%%%%%%%%%%%%%%%%%%%%%%%%%%%%%%%%%%%%%%%%%%%%%%%%%%%%%%%%%%%%

\clearpage

% !TEX root =  2d_3d_kondo_prl_main.tex

\setcounter{figure}{0}
\setcounter{equation}{0}
%\numberwithin{equation}{section}
%\numberwithin{figure}{section}
\renewcommand\thefigure{S\arabic{figure}}
\renewcommand\theequation{S\arabic{equation}}

\title{Supplemental Material for ``Marginal Fermi liquid at magnetic quantum criticality from dimensional confinement''}

\begin{abstract}
\centering
The Supplemental Material contains details on the calculations described in the main text.
\end{abstract}

\date{\today}
\maketitle

%%%%%%%%%%%%%%%%%%%%%%%%%%%%%%%%%%%%%%%%%%%%%%%%%%%%%%%%%%%%%%%%%%%%%%%

\section{Fermion Green's functions}
%
%\lj{If we want section numbers, I believe we either need to change the style of the whole document including the main text (e.g., from PRL to PRB), or make the SM a separate TeX file (with its own style, i.e., PRB). For the submission to PRL, we anyway need to do the latter.}
%
In this section, we detail the structure of the fermion propagators, which form the basis of the effective field theory.
%, and derive Eq.~\eqref{eq:defTmf} of the main text.

\subsection{Conduction-electron Green's function}
\begin{figure}[tb]
\begin{center}
\includegraphics[width=\columnwidth]{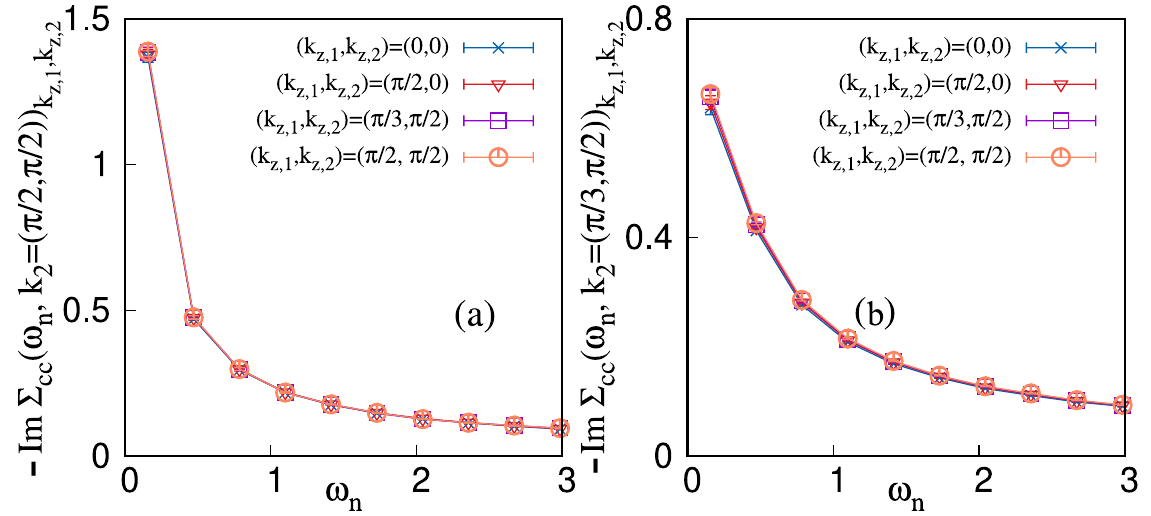}
\caption{Conduction-electron self-energy $\Sigma_{cc}(\omega_n,\boldsymbol{k}_2)_{k_{z,1},k_{z,2}}$ for (a) $\ve k_2 = (\pi/2,\pi/2)$ and (b) $\ve k_2 = (\pi/3,\pi/2)$ as function of frequency $\omega_n$, for different values of $k_{z,1}$ and $k_{z,2}$, demonstrating that $\Sigma_{cc}$ is a constant matrix in the $\{k_{z,1},k_{z,2}\}$ basis. The results are obtained from QMC with $L=12$, $\beta=20$, and $\Jk=3.04 \approx \Jkc$.}
\label{fig:sigma_zz_plot}
\end{center}
\end{figure}

To find the dressed single-particle Green's function $G_{cc}$ of the conduction electrons, we first transform the Kondo interaction of Eq.~\eqref{eq:model} in the main text to momentum space,
\begin{align}\label{eq:ModelFT}
H_{\mathrm K}= \frac{\Jk}{L^2} \sum_{\substack{\ve k_2,k_{z,1}\\ k_{z,2},\ve q_2}} \left(c^\dagger_{\ve  k_2+\ve{q}_2,k_{z,2},\sigma} \boldsymbol{\sigma}_{\sigma\sigma'} c^{}_{\ve  k_2,k_{z,1},\sigma}\right) \cdot \hat{\ve S}^f_{\ve q_2}\, ,
\end{align}
where we have used the standard Fourier representations $c_{\ve j,R_z=0,\sigma} = L^{-3/2} \sum_{\ve k_2,k_z} \exp(i \ve k_2 \cdot \ve j+ i k_z R_z) c_{\ve k_2,k_z,\sigma}$ and $\hat{\ve S}^f_{\ve j} = 1/L \sum_{\ve q_2} \hat{\ve S}^f_{\ve q_2} \exp(i \ve q_2 \cdot \ve j)$.
%Note that the in-plane component of the total momentum is conserved whereas the presence of the spin plane breaks translation invariance in the out-of-plane direction which leads to the violation of momentum conservation in the out-of-plane component.
Since the Kondo interactions act only in the spin layer $R_z=0$, they behave like an effective one-dimensional contact potentials that are constant with respect to $k_z$.
Nevertheless, $G_{cc}(\omega_n,\ve  k_2)_{k_{z,1},k_{z,2}}$ and the corresponding self-energy $\Sigma_{cc}(\omega_n,\ve  k_2)_{k_{z,1},k_{z,2}}$ become non-diagonal matrices in the out-of-plane momentum space. The matrix structure of the latter turns out simple: All contributions to $\Sigma_{cc}$ result from amputated one-particle-irreducible perturbative corrections~\cite{fetter1971} of $G_{cc}$.
In any diagram for  $G_{cc}$, the two external legs with definite $k_{z,1}$ and $k_{z,2}$ are connected via two bare vertices, Eq.~\eqref{eq:ModelFT}, to the rest of the diagram. Since the bare vertex is constant for any choice of the out-of-plane momenta, the dependence on $k_{z,1}$ and $k_{z,2}$ drops out after amputation and, therefore, the self-energy
$\Sigma_{cc}(\omega_n,\ve  k_2)_{k_{z,1}k_{z,2}}$ is independent of both $k_{z,1}$ and $k_{z,2}$. Furthermore, the scaling of $\Sigma_{cc}\sim L^{-1}$ can be derived in analogy to the superficial degree of divergence~\cite{peskin1995}:
Any self-energy diagram of order $(\Jk)^{2n}$ contains $n$ internal spin
propagators $\left\langle \hat{\ve S}^f_{\ve q_2}\hat{\ve S}^f_{\ve q_2} \right \rangle$ and $(2n-1)$ internal bare propagators
$G^{(0)}_{cc}(\omega_n,\ve  k_2)_{k_{z,1}k_{z,2}} \sim \delta_{k_{z,1}k_{z,2}}$. The spin propagators
give rise to momentum sums $(\sum_{\ve q_2})^n \sim L^{2 n}$ while the conduction electrons contribute
% = (i\omega_n-\xi_{\ve  k_2,k_{z,1}})^{-1} \delta_{k_{z,1}k_{z,2}}$. The latter imply summations
$(\sum_{\ve  k_2})^{2n-1} (\sum_{k_z})^{2n-1} \sim L^{2(2n-1)} L^{2n-1}$. Finally, the $2n$ bare vertices provide a factor of $L^{-2\cdot 2n}$.
However, momentum conservation fixes $2n-1$ in-plane momenta and eliminates the corresponding sums.
%are constrained because of momentum conservation at each vertex whereas the out-of plane momenta are not affected at all. Formally, this corresponds to eliminating $2n-1$ sums over $\ve  q_2$ or $\ve  k_2$.
Counting powers of $L$ leads to
\begin{align}
\Sigma_{cc}(\omega_n,\ve  k_2)_{k_{z,1},k_{z,2}} = \frac{\tilde\Sigma_{cc}(\omega_n,\ve  k_2)}{L}\, ,
\end{align}
where $\tilde\Sigma_{cc}$ is both independent of system size and the out-of-plane momenta. The scaling $\Sigma_{cc} \sim L^{-1}$ agrees with the physical picture that interactions take only place in the spin layer, which occupies merely a subsystem of size $L^2$ of the total system with volume $L^3$.
In the matrix-valued Dyson equation for the dressed propagator,
\begin{align}
G_{cc}(\omega_n,\ve k_2)_{k_{z,1} k_{z,2}}^{-1} \!\! =\! G^{(0)}_{cc}(\omega_n,\ve k_2)_{k_{z,1} k_{z,2}}^{-1}\!-\Sigma_{cc}(\omega_n,\ve k_2)_{k_{z,1}k_{z,2}} ,
\label{eq:self-energy-def}
\end{align}
interaction effects appear therefore subleading.
Its solution
\begin{align}\label{eq:GccTmat}
\begin{split}
G_{cc}(\omega_n,\ve k_2 &)_{k_{z,1} k_{z,2}}  =  G^{(0)}_{cc}(\omega_n,\ve k_2,k_{z,1})\delta_{k_{z,1} k_{z,2}}
\\ &  +  G^{(0)}_{cc}(\omega_n,\ve k_2,k_{z,1}) T(\omega_n,\ve k_2) G^{(0)}_{cc}(\omega_n,\ve k_2,k_{z,2})
\end{split}
\end{align}
can be expressed in terms of the $k_z$-independent $T$ matrix
\begin{align}\label{eq:defTgen}
T(\omega_n,\ve k_2) = \frac{1}{L} \dfrac{\tilde \Sigma_{cc}(\omega_n,\ve k_2)}{1-\tilde \Sigma_{cc}(\omega_n,\ve k_2)g_0(\omega_n,\ve k_2)} \equiv \frac{\tilde T(\omega_n,\ve k_2)}{L}
\end{align}
of an effective one-dimensional quantum mechanical scattering potential $\hat V(R_z)= V_0 \delta(R_z)$ with strength $V_0 = \tilde\Sigma_{cc}(\omega_n,\ve k_2)$.
In this sense, the spin plane provides a dynamic scattering barrier, since energy and in-plane momentum are exchanged when the electrons scatter off the spins. Like $\Sigma_{cc}$, also the $T$ matrix is independent of the out-of-plane momenta and scales with $L^{-1}$.
%The local bare propagator in the plane $G^{(0)}_{cc}(\omega_n,\ve  k_2,R_z=0)$ is given below Eq.~\eqref{eq:defTmf} in the main text with the replacement $\omega \to \omega_n$.
%We point out that $\tilde \Sigma_{cc}(\omega_n,\ve  k_2)$ denotes the constant matrix element of $\tilde{\Sigma}_{cc}(\omega_n,\ve  k_2)_{k_{z,1}k_{z_2}}$ which implies that also the $T$-matrix $T(\omega_n,\ve  k_2)_{k_{z,1}k_{z_2}}$ is constant in out-of-plane momentum space with its element given in Eq.~\eqref{eq:defTgen}.
Finally, we Fourier transform both out-of-plane momentum arguments of $G_{cc}(\omega_n,\ve k_2)_{k_{z,1} k_{z,2}}$ to $R_z=0$ and obtain the dressed local propagator
\begin{align}\label{eq:defgfull}
g(\omega_n,\ve k_2) = g_0(\omega_n,\ve k_2) + g_0(\omega_n,\ve k_2)\tilde T(\omega_n,\ve k_2)g_0(\omega_n,\ve k_2)\, ,
\end{align}
from Eq.~\eqref{eq:GccTmat}, which is independent of $L$.
%Consequently, the local counterpart $T(\omega_n,\ve  k_2,R_z=0) = 1/L \sum_{k_{z,1},k_{z_2}} T(\omega_n,\ve  k_2)_{k_{z,1}k_{z,2}}$ reads
%\begin{align}
%T(\omega_n,\ve  k_2,R_z=0) =\dfrac{\tilde \Sigma_{cc}(\omega_n,\ve  k_2)}{1-\tilde \Sigma_{cc}%(\omega_n,\ve  k_2)G^{(0)}_{cc}(\omega_n,\ve  k_2,R_z=0)}\,.
%\end{align}
As presented in Fig.~\ref{fig:sigma_zz_plot}, the QMC results confirm that the self-energy $\Sigma_{cc}(\omega_n,\ve  k_2)_{k_{z,1},k_{z,2}}$ is a constant matrix in the  $\{k_{z,1},k_{z,2}\}$ basis.

\subsection{QMC computation of self-energies}
In the QMC simulations, the self-energy is obtained by computing Eq.~(\ref{eq:self-energy-def}) from the interacting Green's function $G_{cc}(\omega_n,\boldsymbol{k}_2)$. In this study, following the setting of Ref.~\onlinecite{liu2022}, we implement twisted boundary conditions in z direction to reduce the finite size effects. For example, we perform $N_p$ parallel simulations on models with different twist boundary conditions that satisfy $\Phi_{n_p}/\Phi_{0}=1/n_p=\delta_{n_p}$, $n_p\in[0,N_p-1]$. In practice, we perform $N_p>10$ parallel simulations. Observables are determined by averaging over the twist boundary condition. The dressed local propagator defined in $R_z=0$ is given as 
\begin{align}
 & g\left(\omega_{n},\boldsymbol{k}_{2}\right)\nonumber \\
 & =\frac{1}{N_{p}L}\sum_{n_{p},k_{z,1},k_{z,2}}G_{cc}^{\{\Phi_{n_{p}}\}}\left(\omega_{n},\boldsymbol{k}_{2}\right)_{k_{z,1}+\frac{2\pi}{L}\delta_{n_{p}},k_{z,2}+\frac{2\pi}{L}\delta_{n_{p}}} \, ,
\end{align}
where $G_{cc}^{\{\Phi_{n_{p}}\}}\left(\omega_{n},\boldsymbol{k}_{2}\right)_{k_{z,1}+\frac{2\pi}{L}\delta_{n_{p}},k_{z,2}+\frac{2\pi}{L}\delta_{n_{p}}}$ represent the interacting propagator in the presence of the twist boundary conditions $\Phi_{n_p}/\Phi_{0}$. Since the self-energies are related to the inverse propagator, the latter are only defined with respect to a given twist boundary condition. According to the above principle, we obtain the physical self-energies by taking the average $\tilde{\Sigma}_{cc}\left(\omega_{n},\boldsymbol{k}_{2}\right)=\frac{1}{N_p}\sum_{n_p}\tilde{\Sigma}_{cc}^{\{\Phi_{n_p}\}}\left(\omega_{n},\boldsymbol{k}_{2}\right)$, where
\begin{align}
 & \tilde{\Sigma}_{cc}^{\{\Phi_{n_{p}}\}}\left(\boldsymbol{k}_{2},\omega_{n}\right)=\nonumber \\
 & \frac{1}{L}\sum_{k_{z,1},k_{z,2}}\left[\left(G_{cc}^{(0)}\left(\omega_{n},\boldsymbol{k}_{2}\right)^{-1}\right)_{k_{z,1}+\frac{2\pi\delta_{n_{p}}}{L},k_{z,2}+\frac{2\pi\delta_{n_{p}}}{L}}-\right.\nonumber \\
 & \left.\left(G_{cc}^{\{\Phi_{n_{p}}\}}\left(\omega_{n},\boldsymbol{k}_{2}\right)^{-1}\right)_{k_{z,1}+\frac{2\pi\delta_{n_{p}}}{L},k_{z,2}+\frac{2\pi\delta_{n_{p}}}{L}}\right]
\end{align}
are the self-energies for fixed twist boundary condition.

\subsection{Composite-fermion Green's function}
\begin{figure}[tb]
\begin{center}
\includegraphics[width=\columnwidth]{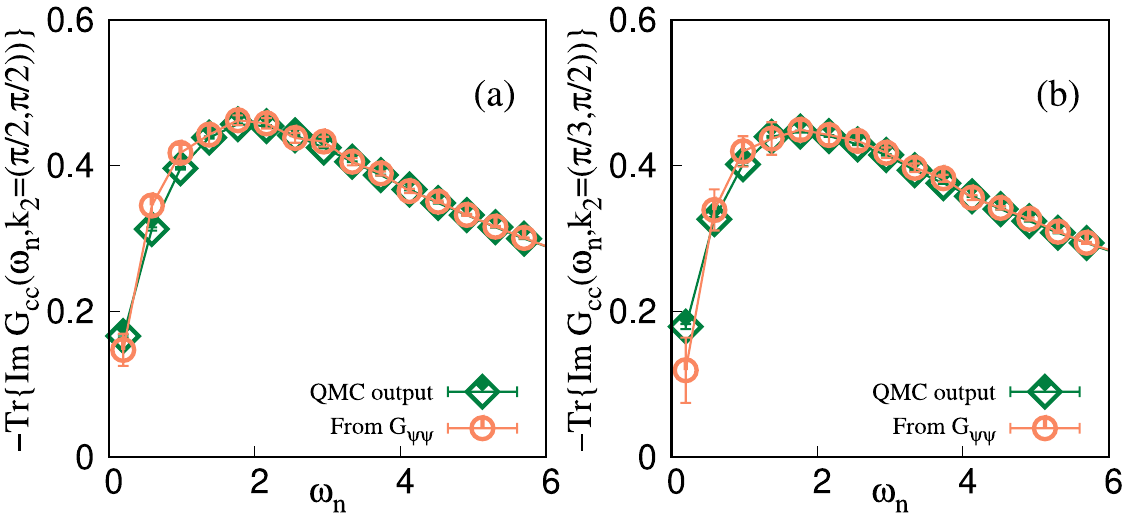}
\caption{Imaginary part of Green's function $G_{cc}$ as function of frequency $\omega_n$ from QMC using $L=12$, $\beta=16$, and $\Jk=3.04 \approx \Jkc$. Trace denotes sum over spin indices. Green squares denote direct QMC output. Orange dots are obtained by employing Eq.~(\ref{eq:GccGpsi}), using the propagator $G_{\psi\psi}$ from QMC.}
\label{fig:Gcc_compare}
\end{center}
\end{figure}

Next, we establish the connection between the propagator of the composite fermions and the $T$ matrix, Eq.~\eqref{eq:defTmf} in the main text. We proceed along the lines of Ref.~\cite{borda2007}, but have to keep track of the dimensional mismatch. Consider the generating functional for correlation functions of the conduction electrons
\begin{align}\label{eq:GenFun}
\mathcal{Z}[\eta,\bar \eta] = \int \mathcal{D}[\ve S^f] \mathcal D[\bar c,c] e^{-S[\ve S^f,c,\bar c] +S_{\text{source}}[c,\bar c, \eta,\bar \eta]} \, ,
\end{align}
expressed as path integral over the Grassmann fields $c_{\ve k_2,k_z,\sigma}(\tau),\bar c_{\ve k_2,k_z,\sigma}(\tau) $ and the variables $\ve S^f_{\ve q_2}(\tau)$ characterizing spin-coherent states~\cite{altland2010} in imaginary time $\tau \in [0,\beta)$. With the short-hand notation  $\int_{\tau_1,...,\tau_n} (\cdot) = \int_0^\beta d\tau_1 ... d\tau_n (\cdot)$, the action $S$ for the Kondo heterostructure [Eq.~\eqref{eq:model} in the main text] reads
\begin{align}\label{eq:Sorig}
S[\ve S^f,c,\bar c]  = S_0[c,\bar c] + S_\mathrm{K}[\ve S^f, c,\bar c] +S_\mathrm{H}[\ve S^f],
\end{align}
with
\begin{align}
S_0  =-\!\! \int_{\substack{\tau \\ \tau'}} \sum_{\substack{\ve k_2,k_z,\\ \sigma}} \!\!\!  \bar c_{\ve k_2,k_z,\sigma}(\tau)
G_{cc}^{(0)}(\tau-\tau',\ve k_2,k_z)^{-1} c_{\ve k_2,k_z,\sigma}(\tau')
\end{align}
and
\begin{align}
S_\mathrm{K}  = \frac{\Jk}{2L^2}\int_\tau  \sum_{\substack{\ve k_2,k_z,k_z'\\ \ve q_2,\sigma,\sigma'}} \!\!\!\!\bar c_{\ve k_2+\ve q_2,k_z,\sigma}(\tau) \boldsymbol \sigma_{\sigma \sigma'}  c_{\ve k_2,k'_z,\sigma}(\tau) \cdot \ve S^f_{\ve q_2}(\tau)\, ,
\end{align}
and we omit the precise form of the Heisenberg action $S_\mathrm{H}$ since it does not play a role in the following manipulations. The inverse bare Green's function in imaginary time is given by $G_{cc}^{(0)}(\tau-\tau',\ve k_2,k_z)^{-1} = \delta(\tau-\tau') (-\partial_\tau - \epsilon_{\ve k_2,k_z})$.
We use the standard source term in Eq.~\eqref{eq:GenFun} with the Grassmann fields $\eta$ and $\bar \eta$:
\begin{align}\label{eq:Ssource}
\begin{split}
& S_{\text{source}}[c,\bar c, \eta,\bar \eta] = \\
& \!\int_\tau\! \sum_{\ve k_2,k_z,\sigma}( \bar c_{\ve k_2,k_z,\sigma}(\tau)  \eta_{\ve k_2,k_z,\sigma}(\tau) + \bar \eta_{\ve k_2,k_z,\sigma}(\tau) c_{\ve k_2,k_z,\sigma}(\tau))  ,
\end{split}
\end{align}
such that the exact single-particle Green's function is given by
\begin{align}\label{eq:FunDerivGcc}
G_{cc}(\tau-\tau',\ve k_2)_{k_{z,1},\sigma; k_{z,2},\sigma'} = \left. \frac{\delta^2 \mathcal Z[\eta,\bar \eta]}{\delta \bar \eta_{\ve k_2,k_{z,1},\sigma} \delta \eta_{\ve k_2,k_{z,2},\sigma'} } \right|_{\eta = 0 = \bar \eta}  \! .
\end{align}
Next, we shift the integration variables of $\mathcal Z[\eta,\bar \eta]$ as follows
\begin{align}\label{eq:intTrafo}
\begin{split}
c_{\ve k_2,k_z,\sigma}(\tau) & \!\to\! c_{\ve k_2,k_z,\sigma}(\tau)\! -\!\!\! \int_{\tau'}\!\!\! G_{cc}^{(0)}(\tau-\tau',\ve k_2,k_z) \eta_{\ve k_2,k_z,\sigma}(\tau'), \\
\bar c_{\ve k_2,k_z,\sigma}(\tau) & \!\to\! \bar  c_{\ve k_2,k_z,\sigma}(\tau)\! -\!\!\! \int_{\tau'}\!\!\! \bar \eta_{\ve k_2,k_z,\sigma}(\tau') G_{cc}^{(0)}(\tau'-\tau,\ve k_2,k_z) ,
\end{split}
\end{align}
while leaving the spin variables $\ve S^f$ unchanged. Thereby, the measure $\mathcal{D}[\bar c, c]$ is not altered, but new terms, both linear and quadratic in the source fields $\eta$ and $\bar\eta$, are introduced in the exponent,
\begin{widetext}
\begin{align}
%\begin{split}
& -S[\ve S^f,c,\bar c]+S_{\text{source}}[c,\bar c,\eta ,\bar \eta] \to  -S[\ve S^f,c,\bar c]- \int_0^\beta d\tau d\tau' \sum_{\ve k_2,k_z,\sigma} \bar\eta_{\ve k_2,k_z,\sigma}(\tau)G_{cc}^{(0)}(\tau-\tau',\ve k_2,k_z) \eta_{\ve k_2,k_z,\sigma}(\tau') \nonumber \\
&\! + \! \frac{\Jk}{2L^2}\!\!\! \int_{\tau,\tau'} \sum_{\substack{\ve k_2,k_z,k'_z\\ \sigma,\sigma',\ve q_2}}\!\! \left[ \bar c_{\ve k_2+\ve q_2,k_z,\sigma}(\tau) G_{cc}^{(0)}(\tau-\tau',\ve k_2,k'_z)\eta_{\ve k_2,k'_z,\sigma'}(\tau') \!+\!
\bar \eta_{\ve k_2,k_z,\sigma}(\tau' ) G_{cc}^{(0)}(\tau'-\tau,\ve k_2,k_z) c_{\ve k_2-\ve q_2,k'_z,\sigma'}(\tau) \!  \right]\! \boldsymbol \sigma_{\sigma \sigma'}\! \cdot \! \ve S^f_{\ve q_2}\!(\tau)  \nonumber \\
&\!- \frac{\Jk}{2L^2} \int_{\tau,\tau', \tau''}\sum_{\substack{\ve k_2, \ve q_2,k_z \\
k'_z,\sigma,\sigma'}} \bar \eta_{\ve k_2+\ve q_2,k_z,\sigma}(\tau' ) G_{cc}^{(0)}(\tau'-\tau,\ve k_2 +\ve q_2 ,k_z) \boldsymbol \sigma_{\sigma \sigma'}\cdot \! \ve S^f_{\ve q_2}\!(\tau) G_{cc}^{(0)}(\tau-\tau'',\ve k_2 ,k'_z) \eta_{\ve k_2,k'_z,\sigma'}(\tau'') \, .
%\end{split}
\end{align}
%\end{widetext}
Note that, by virtue of the identity $\int_0^\beta d\tau'' G_{cc}^{(0)}(\tau-\tau'',\ve k_2,k_z)^{-1} G_{cc}^{(0)}(\tau''-\tau',\ve k_2,k_z) = \delta(\tau-\tau'')$,  the original source term from Eq.~\eqref{eq:Ssource} cancels identically with the new terms of linear order in $\eta$ and $\bar \eta$ arising from the transformation of $S_0$ in Eq.~\eqref{eq:Sorig}. To make contact with the propagator of the composite fermions $G_{\psi \psi}$, we replace the operators $\psi^\dagger_\sigma$ and $\psi_\sigma$ defined in the main text by Grassmann fields and express them in momentum space,
\begin{align}
\begin{split}
\bar \psi_{\ve k_2,\sigma}(\tau) & = \frac{1}{L^{3/2}} \sum_{k_z,\sigma,\ve q_2} \bar c_{\ve k_2+\ve q_2,k_z,\sigma'}(\tau) \boldsymbol \sigma_{\sigma' \sigma} \cdot \ve{S}^f_{\ve q_2}(\tau)\,, \\
\psi_{\ve k_2,\sigma}(\tau) & = \frac{1}{L^{3/2}} \sum_{k_z,\sigma,\ve q_2} \ve{S}^f_{\ve q_2}(\tau) \cdot \boldsymbol \sigma_{\sigma \sigma'} c_{\ve k_2-\ve q_2,k_z,\sigma'}(\tau) \, .
\end{split}
\end{align}
Rewriting the action in terms of the composite fermions yields then
%\begin{widetext}
\begin{align}
%\begin{split}
& -S[\ve S^f,c,\bar c]+S_{\text{source}}[c,\bar c,\eta ,\bar \eta] \to  -S[\ve S,c,\bar c]- \int_0^\beta d\tau d\tau'\sum_{\ve k_2,k_z,\sigma} \bar\eta_{\ve k_2,k_z,\sigma}(\tau)G_{cc}^{(0)}(\tau-\tau',\ve k_2,k_z) \eta_{\ve k_2,k_z,\sigma}(\tau') \nonumber \\
& +  \frac{\Jk}{2L^{1/2}} \int_{\tau,\tau'} \sum_{\substack{\ve k_2,k_z,k'_z\\ \sigma,\sigma',\ve q_2}}\!\! \left[ \bar \psi_{\ve k_2,\sigma}(\tau) G_{cc}^{(0)}(\tau-\tau',\ve k_2,k_z)\eta_{\ve k_2,k_z,\sigma}(\tau') +
\bar \eta_{\ve k_2,k_z,\sigma}(\tau' ) G_{cc}^{(0)}(\tau-\tau',\ve k_2,k_z) \psi_{\ve k_2,\sigma}(\tau')  \right] \nonumber \\
&- \frac{\Jk}{2L^2} \int_{\tau,\tau, \tau''} \sum_{\substack{\ve k_2, \ve q_2,k_z  \\
k'_z,\sigma,\sigma'}} \bar \eta_{\ve k_2+\ve q_2,k_z,\sigma}(\tau ) G_{cc}^{(0)}(\tau-\tau',\ve k_2 +\ve q_2 ,k_z) \boldsymbol \sigma_{\sigma \sigma'}\cdot \! \ve S^f_{\ve q_2}\!(\tau') G_{cc}^{(0)}(\tau'-\tau'',\ve k_2 ,k'_z) \eta_{\ve k_2,k'_z,\sigma'}(\tau'') \, .
%\end{split}
\end{align}
\end{widetext}
Since the transformation of integration variables in Eq.~\eqref{eq:intTrafo} does not change the generating functional $\mathcal{Z}[\eta,\bar \eta]$ we can still apply Eq.~\eqref{eq:FunDerivGcc} to obtain $G_{cc}$, which becomes, as function of Matsubara frequency $\omega_n$,
\begin{align}\label{eq:GccGpsi}
\begin{split}
& G_{cc}(\omega_n,\ve k_2)_{\substack{ k_{z,1} \sigma \\ k_{z,2} \sigma'}} =
G^{(0)}_{cc}(\omega_n,\ve k_2,k_{z,1}) \delta_{k_{z,1},k_{z,2}} \delta_{\sigma \sigma'} \\
& + \frac{\Jk^2}{4L} G^{(0)}_{cc}(\omega_n,\ve k_2,k_{z,1}) G_{\psi \psi}(\omega_n,\ve k_2) G^{(0)}_{cc}(\omega_n,\ve k_2,k_{z,2}) \\
& + \frac{\Jk}{2 L^2} G^{(0)}_{cc}(\omega_n,\ve k_2,k_{z,1}) \boldsymbol \sigma_{\sigma,\sigma'} \cdot \left \< \ve S^f_{\substack{\ve q_2=0\\ \Omega_m=0}} \right\> G^{(0)}_{cc}(\omega_n,\ve k_2,k_{z,2}) \, .
\end{split}
\end{align}
Here, the propagator of the composite fermions is defined as $G_{\psi\psi}(\tau-\tau',\ve k_2)_{\sigma,\sigma'}= -\left\<\mathcal T_\tau[ \psi_{\ve k_2\sigma}(\tau)  \bar\psi_{\ve k_2\sigma'}(\tau')] \right \>$, where $\mathcal T_\tau$ denotes time ordering. In the paramagnetic heavy-fermion phase, we have $G_{\psi \psi}(\omega_n,\ve k_2)_{\sigma \sigma'} \sim \delta_{\sigma \sigma'}$, since, in the absence of magnetic order, $\psi_\sigma$ transforms like a SU(2) spinor (i.e., like conduction electrons) under global spin rotations~\cite{danu21}.
In the above equation, the last line takes the interaction with a finite, static uniform magnetization $\<\ve S^f_{\ve q_2 =0}(\Omega_m=0)\>$ into account. Since the Kondo heterostructure with antiferromagnetic exchange couplings $\Jh,\Jk > 0$ never orders ferromagnetically, the last line vanishes and we obtain
\begin{align}
G_{\psi \psi}(\omega_n,\ve k_2)\! = \!\frac{4}{\Jk^2} \tilde T(\omega_n,\ve k_2)
\end{align}
%\frac{4 \Jk^{-2}}{\dfrac{i \omega_n - \tilde\epsilon_{\ve k_2}}{\vmf^2}-g_0(\omega_n,\ve k_2)} 
%This general result confirms the first part of Eq.~\eqref{eq:defTmf} from the main text, which is evaluated at the mean-field level.
We emphasize that the above result holds generally, beyond the level of mean-field theory. However, applying it to the mean-field calculation confirms the first part of Eq.~\eqref{eq:defTmf} in the main text.
Figure~\ref{fig:Gcc_compare} demonstrates that the QMC data are consistent with Eq.~(\ref{eq:GccGpsi}).

\section{Mean-field $\boldsymbol T$ matrix}
In this section, we derive the mean-field form of the $T$ matrix, introduced in Eq.~\eqref{eq:defTmf} in the main text, and confirm, on the basis of the QMC data, that it indeed serves as suitable starting point of the diagrammatic analysis. Reference~\cite{liu2022} provides an in-depth analysis of the Kondo heterostructure on the mean-field level, which we use here as basis. First, we represent the spin operators in terms of Abrikosov pseudo fermions $\hat{\boldsymbol{S}}_{\ve{i}}^{f}=\frac{1}{2}\ve{\hat{f}}^{\dagger}_{\ve{i}}\boldsymbol{\sigma}\ve{\hat{f}}^{\dagger}_{\ve{i}}$, with $\ve{\hat f}_{\ve i}=(\hat f_{\ve i, \uparrow},f_{\ve i,\downarrow})$. This formulation requires the additional constraint $\hat{Q}_{\ve{i}}= \ve{\hat{f}}^{\dagger}_{\ve{i}} \ve{\hat{f}}^{\phantom\dagger}_{\ve{i}} =1 $, ensuring half filling. In this way, mean-field theory is transformed into an $(L+1)$-site problem for any given $\ve k_2$, corresponding to $L$ conduction-band sites and one $f$ site. In particular, the paramagnetic heavy-fermion phase
is characterized by a finite hybridization parameter $V=\left\langle \hat{c}_{\ve{i},R_z=0,\sigma}^{\dagger}\hat{f}_{\ve{i}, \sigma}^{\phantom\dagger}+\hat{f}_{\ve{i}, -\sigma}^{\dagger}\hat{c}_{\ve{i}, R_z=0,-\sigma}^{\phantom\dagger}\right\rangle \neq 0$
whereas the antiferromagnetic order, without loss of generality considered only along the $z$ direction, vanishes, $\left\langle \hat{S}_{\ve{i}, R_{z}=0}^{c,z}\right\rangle =-m_{c}e^{i\boldsymbol{Q} \cdot\boldsymbol{i}} = 0$ and $\left\langle \hat{S}_{\ve{i}}^{f,z}\right\rangle =m_{f}e^{i\boldsymbol{Q} \cdot\boldsymbol{i}} = 0$. 
Within mean-field theory, the local constraint $\hat Q_{\ve i}$ is implemented only on average via an additional Lagrange parameter $\lambda$. Particle-hole symmetry dictates $\lambda = 0$. As a result, one finds an enhanced Luttinger count $n_c+1$ from the contribution of the pseudo fermions via the hybridization with the conduction electrons~\cite{liu2022}.

\begin{figure}[tb]
\includegraphics[width=\columnwidth]{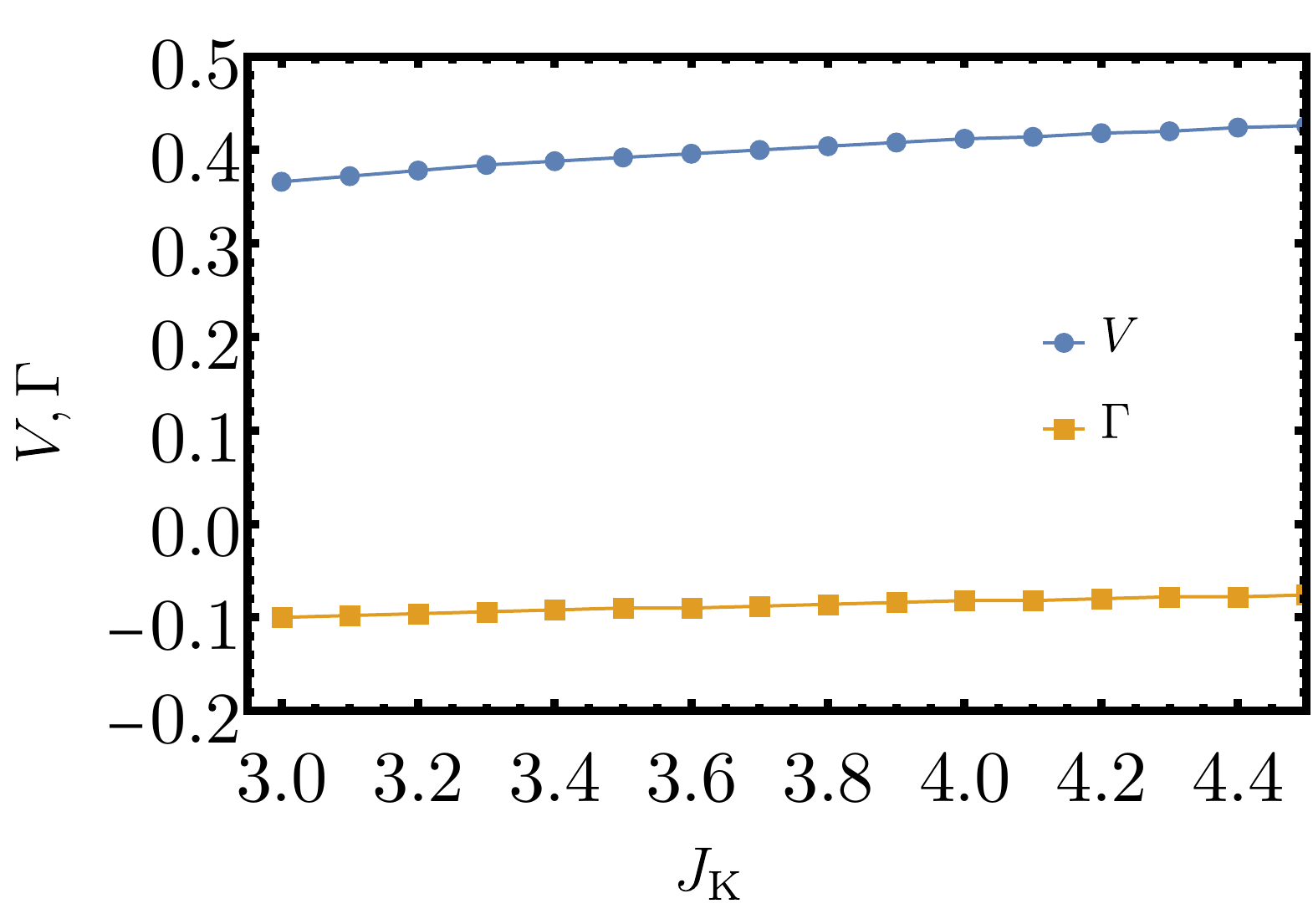}
\caption{Mean-field solution for hybridization parameter $V$ and hopping amplitude $\Gamma$ as function of $\Jk > \Jkc \approx 3.04 t$, assuming a paramagnetic ground state~\cite{liu2022}.}
\label{fig:mfPlot}
\end{figure}

\begin{figure}[t]
\includegraphics[width=\columnwidth]{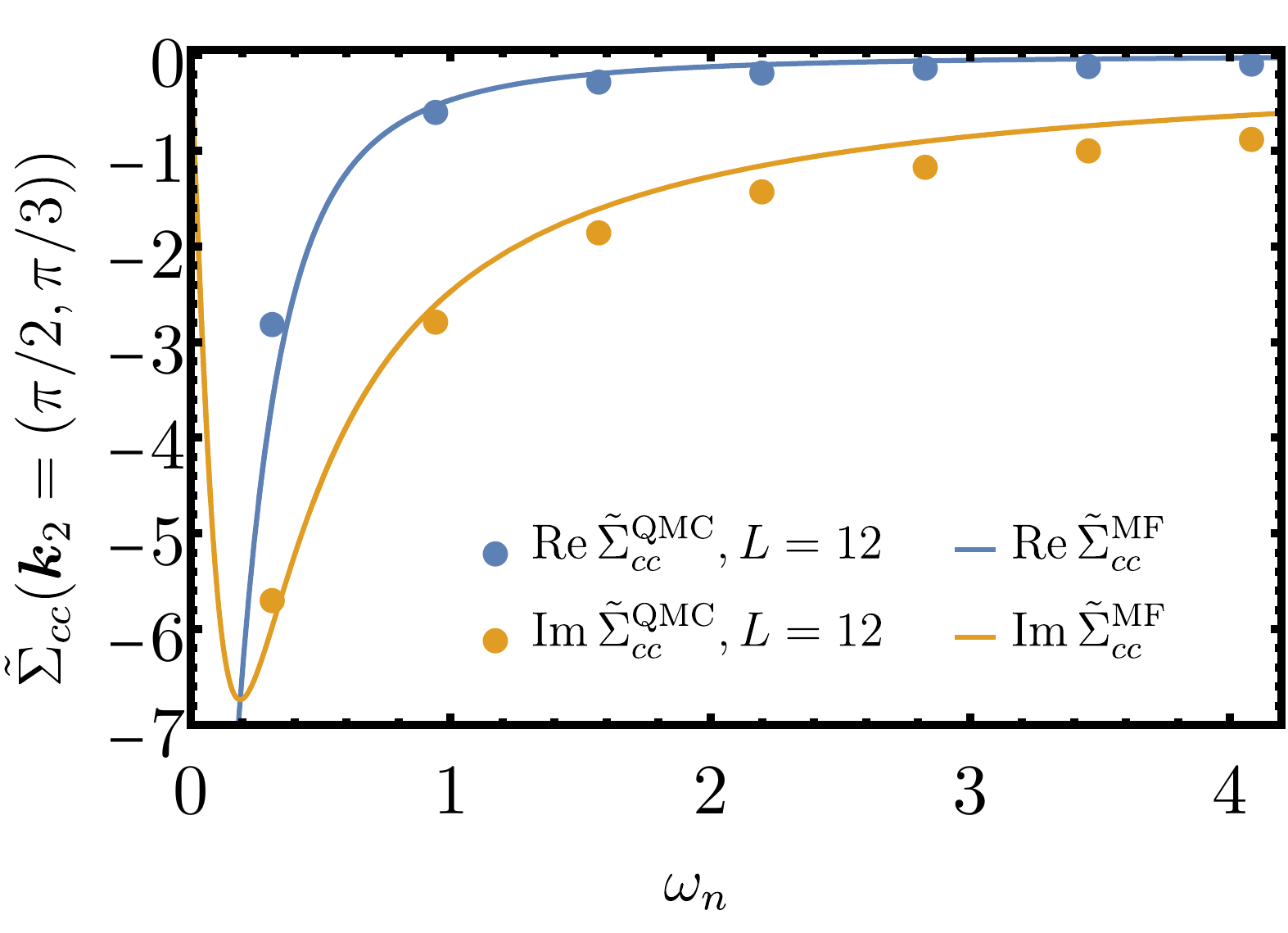}
\caption{Real (blue) and imaginary (ocher) part of conduction-electron self-energy at in-plane momentum $\ve k_2 = (\pi/2,\pi/3)$ and $T/t = 1/10$ from QMC for $\Jk = 3.04 t \approx \Jkc$ (dots) and renormalized mean-field theory (solid lines), $\tilde \Sigma_{cc}^\mf = (i\omega/v^{2}_\mf - t'/t \, \epsilon_{\ve k_2})^{-1}$ [Eq.~\eqref{eq:SigmaccMF}], with $t'/t = -0.08$ and $\vmf = 1.6t$.}
%Comparison of the QMC data for the local self-energy of the conduction electrons $\tilde \Sigma_{cc}$ with the renormalized form $\tilde \Sigma_{cc} = (i\omega/v^{2}_\mf - t'/t \epsilon_{\ve k_2})^{-1}$ of the mean-field result~\eqref{eq:SigmaccMF} at $T=1/10$ and in-plane momentum $\ve k_2 = (\pi/2,\pi/3)$.
\label{fig:KondRes1}
\end{figure}

\begin{figure}[t]
\includegraphics[width=\columnwidth]{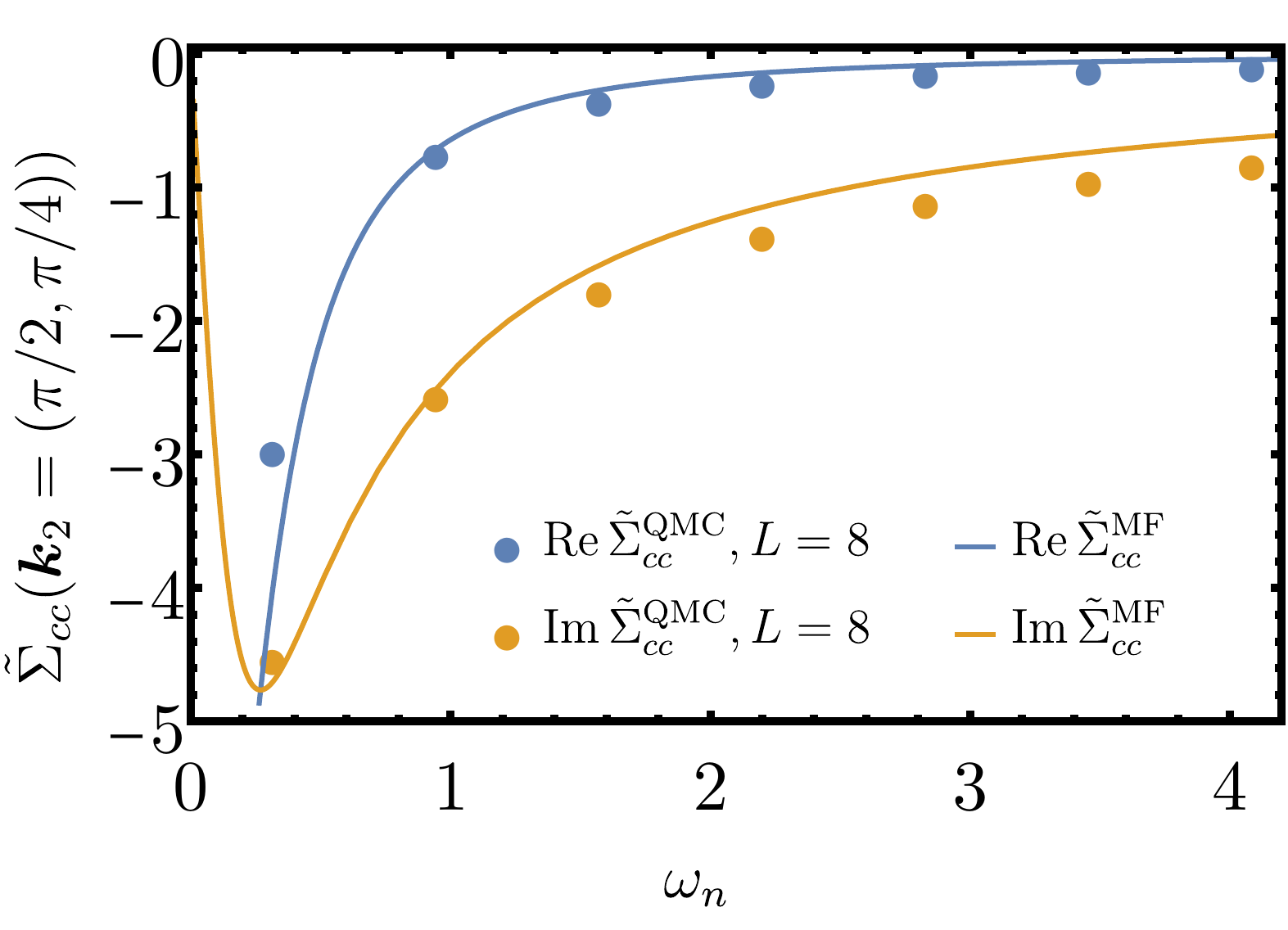}
\caption{Same as Fig.~\ref{fig:KondRes1}, but at in-plane momentum $\ve k_2 = (\pi/2,\pi/4)$.}
%Comparison of the QMC data for the local self-energy of the conduction electrons $\tilde \Sigma_{cc}$ with the renormalized form $\tilde \Sigma_{cc} = (i  \omega/v^{2}_\mf - t'/t \epsilon_{\ve k_2})^{-1}$ of the mean-field result~\eqref{eq:SigmaccMF} at $T=1/10$ and in-plane momentum $\ve k_2 = (\pi/2,\pi/4)$.
\label{fig:KondRes2}
\end{figure}

To make contact with $\tilde T_\mf$, we extend the mean-field Hamiltonian for the paramagnetic phase [Eq.~(D1) of Ref.~\cite{liu2022}] by introducing the hopping parameter $\Gamma$, in analogy to Ref.~\cite{Iglesias1997},
\begin{align}
\Gamma = \left \langle f^\dagger_{\ve i, \uparrow} f_{\ve i+\boldsymbol \delta, \uparrow}  + f^\dagger_{\ve i+\boldsymbol \delta, \downarrow} f_{\ve i, \downarrow}\right \rangle = \left \langle f^\dagger_{\ve i, \downarrow} f_{\ve i+\boldsymbol \delta, \downarrow}  + f^\dagger_{\ve i+\boldsymbol \delta, \uparrow} f_{\ve i, \uparrow}\right \rangle \, .
\end{align}
Formally, it can be obtained by decoupling the $S^{f,x}_{\ve i} S^{f,x}_{\ve i+\delta}+S^{f,y}_{\ve i} S^{f,y}_{\ve i+\delta}$ terms of the nearest-neighbor Heisenberg Hamiltonian [Eq.~\eqref{eq:model} of the main text] after introducing the $f$ fermions. Physically, it describes how the pseudo fermions acquire a dispersion at finite $V$ via hopping processes through the conduction band. For simplicity, we have assumed that $\Gamma$ is independent of the lattice site.
Upon Fourier transforming to in-plane momentum space, the extended mean-field Hamiltonian reads
\begin{align}\label{eq:HMFapp}
\hat{H}_{\text{MF}} & = \!\!\!\!\sum_{\substack{\ve k_2, R_z\\ \sigma}}\!\!\! \left[\epsilon_{\ve k_2} c^\dagger_{\ve k_2, R_z,\sigma} c_{\ve k_2, R_z,\sigma}\!-\!t(c^\dagger_{\ve k_2, R_z+1,\sigma} c_{\ve k_2, R_z,\sigma}\! + \!\text{h.c.})\!\right] \nonumber \\
& -\frac{\Jk V}{2}\sum_{\ve k_2,\sigma}\left(\hat{c}_{\ve k_2,R_z=0,\sigma}^{\dagger}\hat{f}_{\ve k_2,\sigma}+\text{h.c.}\right)
+\frac{L^2 \Jk V^2}{2} \nonumber \\
&+\frac{\Jh \Gamma}{4 t} \sum_{\ve k_2\sigma}
\epsilon_{\ve k_2} f^\dagger_{\ve k_2, \sigma} f_{\ve k_2, \sigma} + \frac{\Jh z_\text{c} L^2 \Gamma^2}{4} \, .
\end{align}
Here, $z_\text{c}=4$ is the coordination number of the square lattice and $\epsilon_{\ve k_2}= -2t (\cos k_x + \cos k_y)$. In the ground state, the mean-field parameters satisfy $\partial\<\hat H_\mf\>/\partial V= 0$ and $\partial\<\hat H_\mf\>/\partial \Gamma= 0$. The numerical solution is presented in Fig.~\ref{fig:mfPlot}. The ground-state energy can be decreased by taking a small, negative value of $\Gamma$, which means that the pseudo fermions acquire an inverted nearest-neighbor tight-binding dispersion. The interactions effects at the mean-field level can be read off from the hybridization part of $\hat H_\mf$ from Eq.~\eqref{eq:HMFapp} in momentum space, $-\Jk V/(2 L^{1/2})\cdot\sum_{\ve k_2,k_z,\sigma}\left(\hat{c}_{\ve k_2,k_z,\sigma}^{\dagger}\hat{f}_{\ve k_2,\sigma}+\text{h.c.}\right)$, implying that the conduction-electron self-energy has the mean-field form
\begin{align}\label{eq:SigmaccMF}
\Sigma^\mf_{cc}(\omega,\ve k_2)_{k_z, k'_z} = \frac{\Jk^2 V^2}{4L} G^{(0)}_{ff}(\omega, \ve k_2) = \frac{\Jk^2 V^2}{4L} \frac{1}{i \omega - \dfrac{\Jh \Gamma}{4t}\epsilon_{\ve k_2}}\, .
\end{align}
This form describes a Kondo resonance with an inverted dispersion, since $\Gamma < 0$. 
Inserting the self-energy $\Sigma^\mf_{cc}$ into the $T$ matrix from Eq.~\eqref{eq:defTgen} gives rise to the mean-field $T$ matrix given in Eq.~\eqref{eq:defTmf} of the main text. 
Using the mean-field parameters $V = 0.366$ and $\Gamma = -0.1$ at $\Jk = 3t$, one obtains crude estimates for the hopping amplitude $t' = \Jh \Gamma/(\Jk^2 V^2) \approx -0.04 t$ and the weight of the Kondo resonance $v_\mf = \Jk V/2 \approx 0.55 t$. 
Comparison with the QMC data at $T/t=1/10$ for $\Sigma_{cc}$ at the coupling $\Jk=3.04 t \approx \Jkc$ indeed reveals the dispersing scattering pole, as exemplified in Figs.~\ref{fig:KondRes1} and~\ref{fig:KondRes2}.
%Hence, the comparison to the mean-field theory at $T=0$ is justified, while we can evaluate the mean-field self-energy at finite temperatures by inserting the fermionic Matsubara frequencies, i.e., $\Sigma_{cc}^\mf(i\omega \to i \omega_n)$.
%\lj{$\Sigma_{cc}^\mf(\omega \to i \omega_n)$?}
To capture possible renormalizations of the mean-field values, we replace $\Jh \Gamma/(\Jk^2 V^2) \to t'$ and $\Jk^2 V^2/4 \to v^2_\mf$ in Eq.~\eqref{eq:SigmaccMF} and obtain the values for $t'=-0.08 t$ and $v_\mf=1.6 t$ by fitting to the QMC data, see Sec.~\ref{sec:matching}. Reassuringly, these are in the same ballpark as the above crude estimates.
%\fa{Did  we not  use  $v_{MF}$ throughout the text? } 
%\bof{Check values} 
%by fitting to the QMC data. 
%Using $\Sigma_{cc}^\mf$ and Eq.~\eqref{eq:defTgen}, gives eventually rise to the $T$ matrix at the mean-field level $\tilde T_\mf$ that is stated in Eq.~\eqref{eq:defTmf} in the main text.

\section{One-loop self-energies}

In this section, we evaluate the one-loop self-energies $\Pi(\Omega,\ve q_2)$ and $\Sigma_{\psi\psi}(\omega,\ve k_2)$ at $T=0$ and show how Landau damping and marginal-Fermi-liquid scaling arise. In the first case, it suffices to consider $\delta\Pi(\Omega,\ve Q)=\Pi(\Omega \to 0, \ve Q)-\Pi(\Omega=0,\ve Q)$, because the condition $M^2(T=0)=M_0^2+\Pi(\Omega,\ve Q)=0$ for the QCP absorbs the constant value $\Pi(\Omega=0,\ve Q)$. Finite momentum deviations from the ordering wavevector turn out to be regular and are not considered in the following.
Using continuous imaginary frequencies, we have at the one-loop level [Fig.~\ref{fig:FeynmanDiag}(a) in the main text],
\begin{align}\label{eq:oneloopPi}
\begin{split}
& \delta\Pi(\Omega \to 0, \ve Q) = \frac{32 \geff^2}{(\Jkc)^4}\int \frac{d^2 k_2}{(2\pi)^2} \int \frac{d\omega}{2\pi} \\
& \times \left[\tilde T_\mf(\omega,\ve k_2) \tilde T_\mf(\omega+\Omega,\ve k_2+\ve Q)- (\tilde T_\mf(\omega,\ve k_2))^2 \right]\, ,
\end{split}
\end{align}
which includes a factor of two from the spin summation.
The dominant contribution in the low-frequency limit is generated by the nonanalytic behavior $g_{0}(\omega \to 0,\ve k_2) \to -i\text{sgn}(\omega)(4t^2-\epsilon_{\ve k_2}^2)^{-1/2}\theta(2t-|\epsilon_{\ve k_2}|)$,
which implies the asymptotics $\tilde T_\mf(\omega \to 0,\ve k_2)=1/[-\tilde \epsilon_{\ve k_2} +i\sgn(\omega)(4t^2-\epsilon^2_{\ve k_2})^{-1/2}]$. Physically, the in-plane momenta for which the step function is finite
correspond to a projection of the 3D Fermi surface into the 2D Brillouin zone, shown in Figs.~\ref{fig:QMCmatch}(a,b) in the main text. The leading behavior of the self-energy in the infrared then reads, using $\epsilon_{\ve k_2+\ve Q} =- \epsilon_{\ve k_2}$,
\begin{align}
%\begin{split}
\delta \Pi(\Omega,\ve Q) & \simeq \frac{32 \geff^2}{(\Jkc)^4}\int_{\substack{\text{proj.} \\ \text{2D FS}}} \frac{d^2 k_2}{(2\pi)^2} \int \frac{d\omega}{2\pi} \nonumber \\
& \qquad\qquad \times \frac{1}{4t^2-\epsilon_{\ve k_2}^2} \frac{1-\sgn(\omega) \sgn(\omega +\Omega)}{\left(\tilde \epsilon_{\ve k_2}^2 +(4t^2 - \epsilon^2_{\ve k_2})^{-1}\right)^2} \nonumber \\
& = \frac{32 \geff^2}{\pi (\Jkc)^4} |\Omega|\! \int_{\substack{\text{proj.} \\ \text{2D FS}}} \! \frac{d^2 k_2}{(2\pi)^2} \frac{(4t^2-\epsilon_{\ve k_2}^2)^{-1}}{\left(\tilde \epsilon_{\ve k_2}^2 +(4t^2 - \epsilon^2_{\ve k_2})^{-1}\right)^2}\, .
%\end{split}
\end{align}
As a result, we obtain indeed the nonanalytic Landau damping term $\delta \Pi(\Omega,\ve Q)\simeq \alpha |\Omega|$ with nonuniversal prefactor, as stated in the main text.

\begin{table*}[tb]
\caption{Summary of all parameters involved in the comparison between effective field theory and QMC.}
\label{numVal}
\begin{tabular*}{\linewidth}{@{\extracolsep{\fill} } l l l}
\hline\hline
Quantity & Value & Obtained via \\
\hline
$\alpha$   & 0.50  & direct fit to $D^{-1}(\Omega_n,\ve Q)-M^2(T)$, see Fig.~\ref{fig:alphaFit}  \\
$d_0   $   & 10.62 & direct fit to $D^{-1}(\Omega_n,\ve Q)-M^2(T)$, see Fig.~\ref{fig:alphaFit}  \\
$M^2(T) = a T \log (b T)$ & $a = - 0.32$, $b= 0.49$ & direct fit to $D^{-1}(\Omega_n=0,\ve Q)$, see Fig.~\ref{fig:QMCmatch}(c) \\
$c_B^2$    & 0.20  &  direct fit to $D^{-1}(\Omega_n=0,\ve q_2)-M^2(T)$, see Fig.~\ref{fig:cbFit} \\ 
$\tilde\epsilon_{\ve k_2} = (t'/t) \epsilon_{\ve k_2}$ & $t'/t= -0.08$ & direct fit to $\Sigma_{cc}$, see Figs.~\ref{fig:KondRes1} and \ref{fig:KondRes2}  \\
$v_\mf$ & $2.58$ & matching with QMC at $T=1/10$ and $\ve k_2 = \ve Q/2$ \\
$g_{\text{eff}}$ & $0.75 \Jk$ & matching with QMC at $T=1/10$ and $\ve k_2 = \ve Q/2$ \\
\hline\hline
\end{tabular*}
\end{table*}

Next, we turn to the one-loop self-energy of the composite fermions [Fig.~\ref{fig:FeynmanDiag}(b) in the main text], which reads at zero temperature
\begin{align}\label{eq:oneloopPsiT0}
\begin{split}
\Sigma_{\psi\psi}(\omega,\ve  k_2)\! =\! \frac{4\geff^2}{(\Jkc)^2}\!\! \int\!\!\!\frac{d^2 q_2}{(2\pi)^2}\!\! \frac{d\Omega}{2\pi}D(\Omega,\ve q_2)\tilde T_\mf(\omega+\Omega,\ve k_2+\ve q_2).
\end{split}
\end{align}
In the above, we have assumed that there is one component of the order parameter field $\ve \Phi$ that shows the critical fluctuations towards the N\'eel-ordered state. Otherwise, one may rescale $g_{\text{eff}}$ accordingly.  Furthermore, we have inserted the dressed antiferromagnetic susceptibility $D(\Omega,\ve q_2)$ at the QCP from Eq.~\eqref{eq:Ddressed}. The infrared singularity of the latter in combination with the discontinuous behavior of $g_0(\omega \to 0, \ve k_2)$ as part of $\tilde T_\mf$ leads to the dominant low-frequency behavior in the projected 2D Fermi surface
\begin{align}\label{eq:MFLT0}
\begin{split}
&\Sigma_{\psi \psi}(\omega \to 0,\ve k_2) \simeq -i\frac{4 \geff^2}{(\Jkc)^2}
 \frac{(4t^2-\epsilon_{\ve k_2}^2)^{-1/2}}{\tilde \epsilon_{\ve k_2}^2 +(4t^2-\epsilon_{\ve k_2}^2)^{-1}} \\
& \times \theta(2t-|\epsilon_{\ve k_2}|)\!\int\! \frac{d\Omega}{2\pi} \int_{|\ve q_2|\leq \Lambda} \! \frac{d^2 q_2}{(2\pi)^2} \frac{1}{c_B^2 q_2^2 +\alpha |\Omega|} \text{sgn}(\Omega+\omega)\, .
 \end{split}
 \end{align}
Here, we have regularized the integral with a high-momentum cutoff $\Lambda$, beyond which the expansion of the momentum dependence becomes inaccurate. 
The ordering wavevector $\ve Q$ has been shifted to the fermionic propagator where it is irrelevant due to $\epsilon_{\ve k_2+ \ve Q} = - \epsilon_{\ve k_2}$. The solution of the momentum integral reads in the limit $|\Omega| \ll c_B \Lambda$
\begin{align}
\begin{split}
&\Sigma_{\psi \psi}(\omega \to 0,\ve k_2) \simeq \frac{-i\geff^2}{\pi (\Jkc)^2 c_B^2}
 \frac{(4t^2-\epsilon_{\ve k_2}^2)^{-1/2}}{ \tilde \epsilon_{\ve k_2}^2 +(4t^2-\epsilon_{\ve k_2}^2)^{-1}} \\
 & \qquad\qquad\times\theta(2t-|\epsilon_{\ve k_2}|) \int \frac{d\Omega}{2\pi} \log \left(\frac{c_B^2 \Lambda^2}{\alpha |\Omega|} \right) \sgn(\Omega+\omega)\, .
\end{split}
\end{align}
Finally, one obtains for the frequency integral
\begin{align}\label{eq:MFL-gs}
\begin{split}
&\Sigma_{\psi \psi}(\omega \to 0,\ve k_2) \simeq \frac{-i\geff^2}{\pi (\Jkc)^2 c_B^2}
 \frac{(4t^2-\epsilon_{\ve k_2}^2)^{-1/2}}{\tilde \epsilon_{\ve k_2}^2 +(4t^2-\epsilon_{\ve k_2}^2)^{-1}} \\
 & \qquad\qquad \quad\theta(2t-|\epsilon_{\ve k_2}|)\times \sgn(\omega) \int_0^{|\omega|} \frac{d\Omega}{\pi} \log \left(\frac{c_B^2 \Lambda^2}{\alpha \Omega} \right) \\
& = \frac{-i\geff^2}{\pi^2 (\Jkc)^2 c_B^2}
 \frac{(4t^2-\epsilon_{\ve k_2}^2)^{-1/2}\, \theta(2t-|\epsilon_{\ve k_2}|)}{\tilde \epsilon_{\ve k_2}^2 +(4t^2-\epsilon_{\ve k_2}^2)^{-1}}  \omega \log \left(\frac{e c_B^2 \Lambda^2}{\alpha |\omega|} \right)  \, .
\end{split}
\end{align}
This result is indeed of the marginal-Fermi-liquid form $\Sigma_{\psi \psi}(\omega \to 0,\ve k_2) = - i \gamma(\varepsilon_{\ve k_2}) \omega \log \left(e c_B^2 \Lambda^2/(\alpha |\omega|)\right)$, given in Eq.~\eqref{eq:resImSigmaT0} of the main text.

\section{Matching with QMC} 
\label{sec:matching}
\begin{figure}[tb]
\begin{center}
\includegraphics[width=\columnwidth]{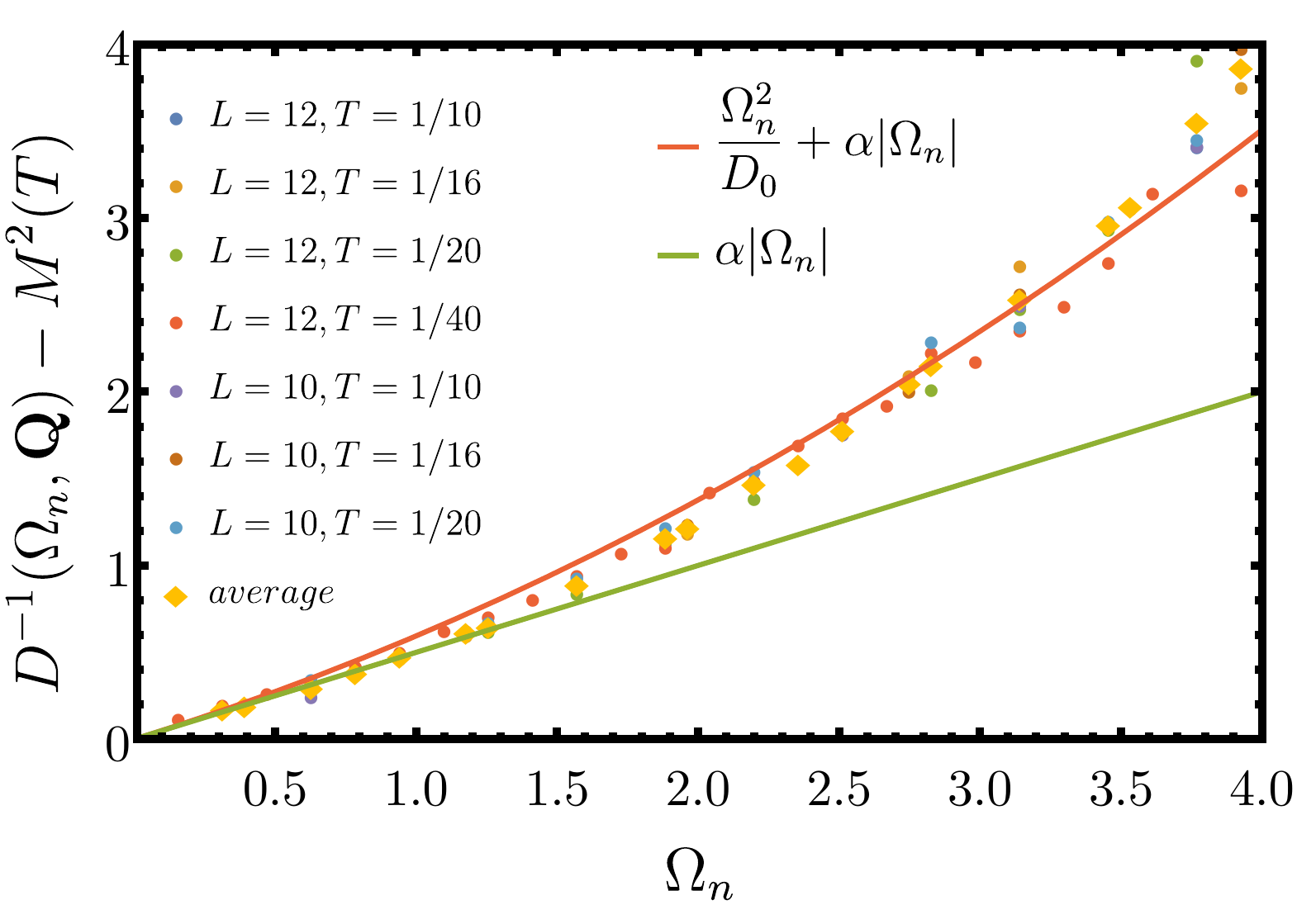}
\caption{Inverse magnetic susceptibility as function of Matsubara frequency $\Omega_n$ at fixed momentum $\ve q_2 = \ve Q$ from QMC simulations for different temperatures $T$ and system sizes $L$. 
Yellow squares: average of the QMC data over system size and temperature. 
Green line: Fit of Landau damping asymptotics $\alpha |\Omega_n|$ to the smallest Matsubara frequencies, with $\alpha = 0.50$.   
Red line: Fit of $d_0^{-1} \Omega_n^2 + 0.5 |\Omega_n|$, with $d_0 = 10.62$, modeling corrections at larger $\Omega_n$.}
\label{fig:alphaFit}
\end{center}
\end{figure}

\begin{figure}[tb]
\begin{center}
\includegraphics[width=\columnwidth]{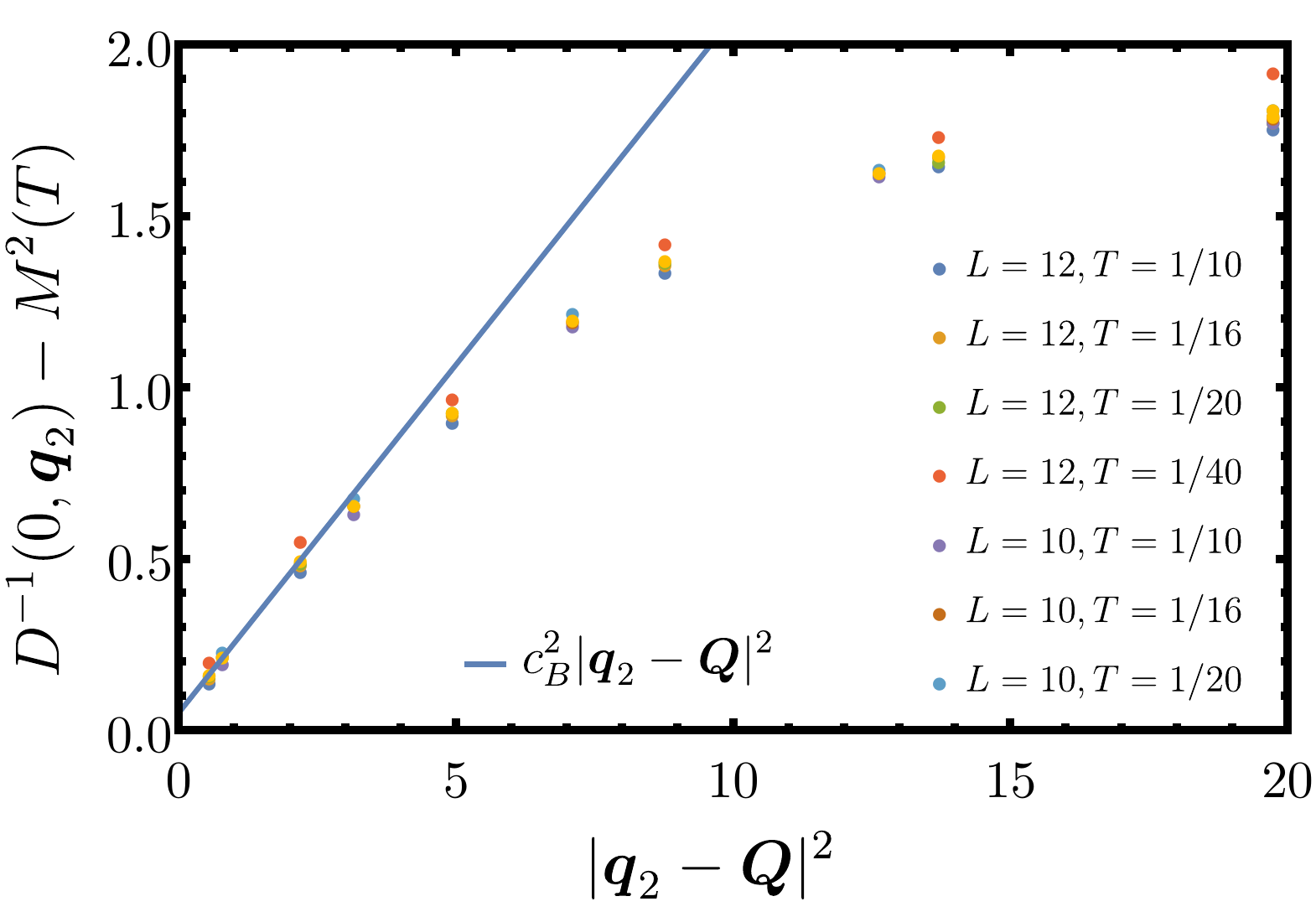}
\caption{Inverse magnetic susceptibility as function of in-plane momentum $\ve q_2$ in vicinity of ordering wavevector $\ve Q$ at fixed Matsubara frequency $\Omega_n=0$ from QMC simulations for different temperatures $T$ and system sizes $L$. Blue line: Fit of asymptotic behavior $c_B^2(\ve q_2-\ve Q)^2$, with $c_B^2 = 0.20$.}
\label{fig:cbFit}
\end{center}
\end{figure}

To identify marginal-Fermi-liquid features at finite temperatures and to compare the effective field theory with the QMC data, we consider the composite-fermion self-energy 
\begin{align}\label{eq:oneloopPsi}
\begin{split}
\Sigma_{\psi\psi}(\omega_n,\ve  k_2) = \frac{4\geff^2}{\Jk^2\beta} \sum_{\omega_m,\ve q_2}\!\!\!D(\omega_m-\omega_n,\ve q_2) \tilde T(\omega_m,\ve k_2+\ve q_2).
\end{split}
\end{align}
at the one-loop level [Fig.~\ref{fig:FeynmanDiag}(b) in the main text], where $\beta=1/T$ corresponds to the inverse temperature. In the above, $D(\Omega_m,\ve q_2)$ refers to the dressed magnetic susceptibility around the ordering wavevector $\ve Q$, as given in Eq.~\eqref{eq:DfiniteT} of the main text, while the $T$ matrix 
%\lj{$T(\omega_n,\ve k_2) = L^{-1} \tilde T(\omega_n,\ve k_2)$ with}
\begin{align}\label{eq:defTsc}
\tilde T(\omega_n,\ve k_2) = \left[\frac{i \omega_n}{v^2_\mf} - \tilde{\epsilon}_{\ve k_2} -\frac{4 \Sigma_{\psi \psi}(\omega_n,\ve k_2)}{\Jk^2} - g_0(\omega_n,\ve k_2)\right]^{-1} \, .
\end{align}
takes the effects of $\Sigma_{\psi\psi}(\omega_n,\ve  k_2)$ self-consistently into account. 
In the following, we express all energies in terms of the bare hopping $t$ and all lengths in terms of the lattice constant $a$, i.e., we set $t=1$ and $a=1$. Furthermore, we consider the coupling strength $\Jk = 3.04 \approx \Jkc$.
Next, we describe how the numerical parameters, necessary for the quantitative comparison between QMC and effective field theory, are obtained in detail. A summary is given in Table~\ref{numVal}. 
First, we fix the parameters of $D$ by fitting the expected low-energy form, Eq.~\eqref{eq:DfiniteT} in the main text, to the QMC data. As shown in Figs.~\ref{fig:alphaFit} and~\ref{fig:cbFit}, the latter asymptotics is indeed confirmed by the QMC simulations. In particular, the linear frequency dependence induced by Landau damping is clearly visible at the smallest frequencies, with prefactor $\alpha = 0.50$.
For the numerical evaluation of the Matsubara sum in Eq.~\eqref{eq:oneloopPsi}, we also include the quadratic frequency variation $d_0^{-1} \Omega^2$, with $d_0=10.62$, important at larger Matsubara frequencies.
In addition, we obtain for the speed of the magnetic modes $c_B^2 = 0.20 $. Note that these parameters do not show a significant temperature dependence.  Finally, we find for the thermal gap parameter $M^2(T) = a T \log(b T)$, with $a=-0.32, b= 0.49$, see Fig.~\ref{fig:QMCmatch}(c) in the main text. 
Next, we turn to the self-consistent $T$ matrix in Eq.~\eqref{eq:defTsc}.
Because we do not observe sizable effects from $\Re\Sigma_{\psi \psi}$, we only take the renormalized mean-field dispersion of the Kondo resonance part $\tilde{\epsilon}_{\ve k_2} = (t'/t) \epsilon_{\ve k_2} = - 0.08 \epsilon_{\ve k_2}$, obtained from the QMC simulations, into account, see Figs.~\ref{fig:KondRes1} and~\ref{fig:KondRes2}. If MFL correlations are present, however, $\Im\Sigma_{\psi \psi}$ has important consequences for the behavior at small Matsubara frequencies but there is no controlled way to decompose the QMC data into the mean-field contribution $i \omega_n / v_\mf^2$  and the beyond-mean field effects contained in $\Im \Sigma_{\psi\psi}(\omega_n)$. As a consequence, using the value $v_\mf=1.6$ from fitting only the mean-field Kondo resonances~\eqref{eq:SigmaccMF} to the QMC self-energies, does not lead to reliable results. Therefore, we keep $v_\mf$, and also the effective Bose-Fermi coupling $\geff$, as two free parameters. These will be fixed by a matching procedure explained below.

For any choice of $v_\mf$ and $\geff$, we compute the Matsubara sum as follows: If the value $D^{-1}(\omega_m-\omega_n,\ve Q) - M^2(T)$ is available from the QMC data for system sizes $L=10$ or $L=12$ and temperatures between $T=1/10$ and $T = 1/40$, we use $D^{-1}(\omega_m-\omega_n,\ve Q) - M^2(T)$ averaged over $L$ and $T$ as input for the frequency-dependent part of the magnetic susceptibility in Eq.~\eqref{eq:oneloopPsi}. This avoids modeling the linear-to-quadratic crossover in $D^{-1}(\omega_m-\omega_n, \ve q_2)$ and reduces the statistical fluctuations.
If $D^{-1}(\omega_m-\omega_n,\ve Q) - M^2(T)$ is not available, i.e., for larger Matsubara frequencies or temperatures below the reach of QMC, we use instead the asymptotics $d_0^{-1} (\omega_m-\omega_n)^2+\alpha |\omega_m-\omega_n|$. Finally, we truncate the Matsubara sum at a maximal frequency $\Omega_n \simeq 20$, which provides stable results in the regime $|\omega_n| \leq 5$. The $\ve q_2$ integrals are taken over the entire 2D Brillouin zone. It turns out that momentum regions away from the ordering wavevector $\ve Q$, where the magnetic susceptibility is no longer given by the asymptotic form, Eq.~\eqref{eq:DfiniteT} in the main text, do not give rise to important contributions if $\ve k_2$ is in the projected 2D Fermi surface. We recall from Eq.~\eqref{eq:MFL-gs} that marginal-Fermi-liquid behavior is anyway only expected for these $\ve k_2$. Self-consistent convergence is in general reached after three to four iterations.   
To determine the only two free parameters $\vmf$ and $\geff$, we match the pseudo-fermion self-energy $\Im \Sigma_{\psi\psi}(\omega_n,\ve k_2)$ as function of $\omega_n$ to the QMC data at fixed in-plane momentum $\ve k_2=(\pi/2,\pi/2)$ and fixed temperature $T=1/10$.  
This yields $\vmf = 2.58$ and $\geff = 0.75 \Jk$.

\begin{figure}[tb]
\begin{center}
\includegraphics[width=\columnwidth]{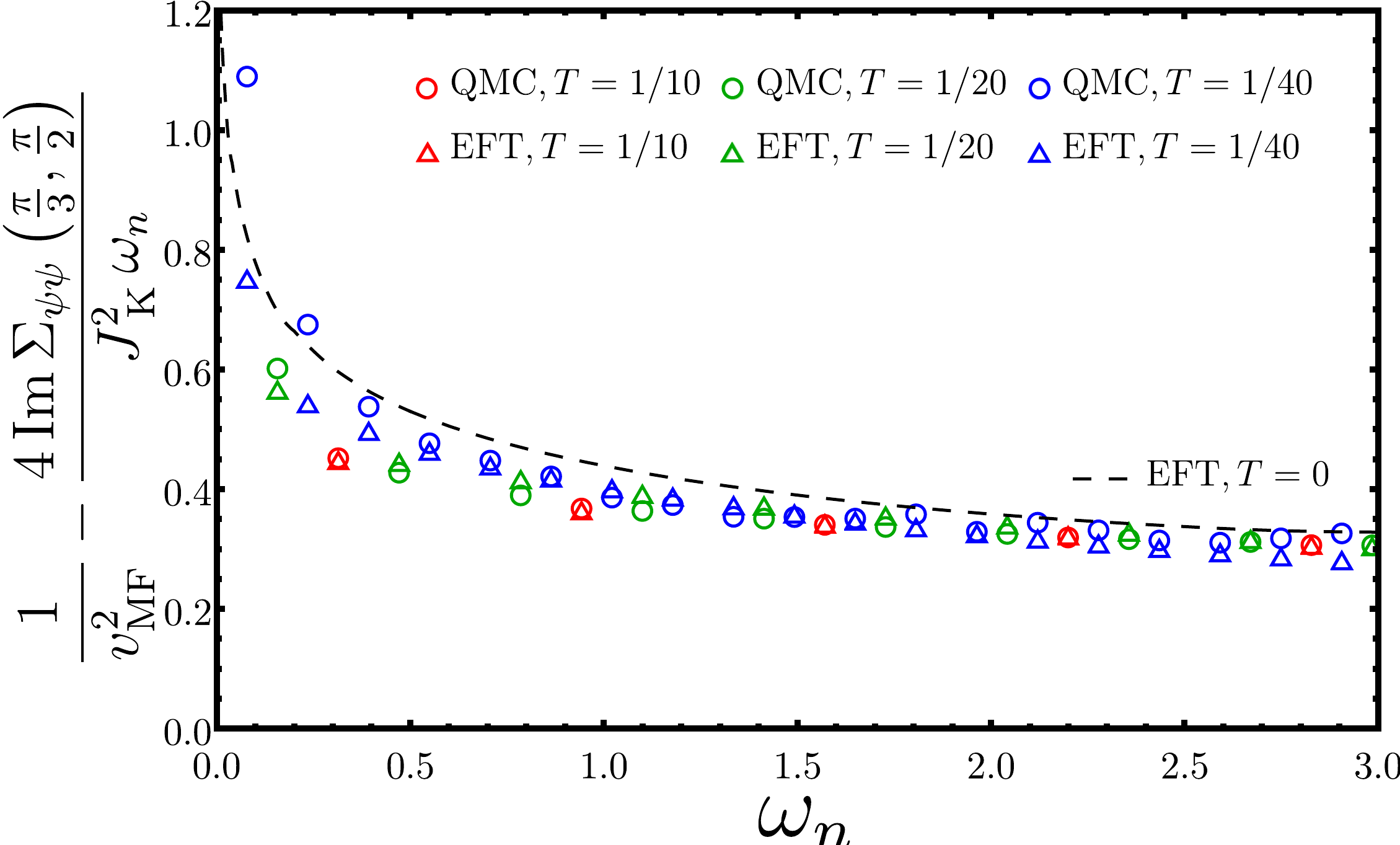}
\caption{Same as Fig.~\ref{fig:QMCmatch}(d) in the main text, but for in-plane momentum $\ve k_2 = (\pi/3,\pi/2)$, inside the projected 2D Fermi surface.
% Comparison of the fermionic self-energy at in-plane momentum $\ve k_2 = (\pi/3,\pi/2)$ from the effective field theory (EFT) with the QMC simulations for system size $L=12$. 
Again, we find a remarkable agreement between effective field theory QMC. The dashed black line shows the extrapolation to $T=0$.}
\label{fig:Sigma105157}
\end{center}
\end{figure}

\begin{figure}[tb]
\begin{center}
\includegraphics[width=\columnwidth]{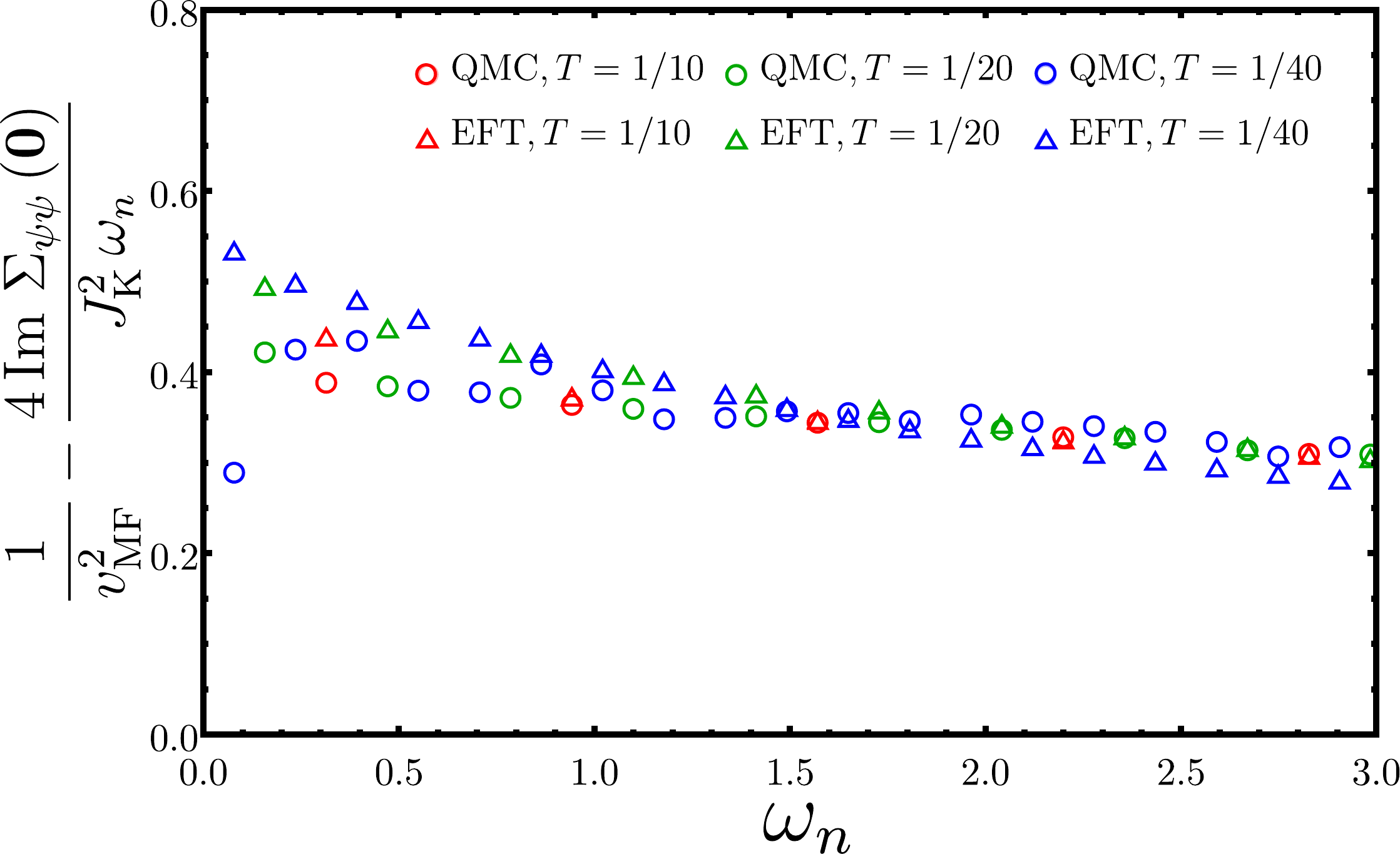}
\caption{Same as Fig.~\ref{fig:QMCmatch}(d) in the main text, but for in-plane momentum $\ve k_2 = \ve \Gamma = (0,0)$.
%Comparison of the fermionic self-energy at the $\Gamma$ point $\ve k_2 = (0,0)$ from the effective field theory (EFT) with the QMC simulations for system size $L=12$. 
As expected, the agreement is reduced, since the $\ve \Gamma$ point is outside the projected 2D Fermi surface, such that $\Sigma_{\psi\psi}$ is not dominated by the combined singular low-energy behavior of $D$ and $T$ in Eq.~\eqref{eq:oneloopPsi}. Consequently, no strong increase of $\Im \Sigma_{\psi\psi}$ at the smallest Matsubara frequencies and no marginal-Fermi-liquid behavior appears, in contrast to the situation for $\ve k_2$ within the projected 2D Fermi surface.}
\label{fig:Sigma000000}
\end{center}
\end{figure}

Having determined all nonuniversal parameters of the effective field theory, we compute next $\Sigma_{\psi\psi}(\omega_n,\ve k_2)$ at lower temperatures and different $\ve k_2$, see Figs.~\ref{fig:Sigma105157} and \ref{fig:Sigma000000}, as well as Fig.~\ref{fig:QMCmatch}(d) in the main text. Provided that $\ve k_2$ is inside of the projected 2D Fermi surface, we obtain for all values of $\ve k_2$ the same level of quantitative agreement. For values of $\ve k_2$ outside the projected 2D Fermi surface, the self-energy is not dominated by the infrared singularities of the magnetic susceptibility and the $T$ matrix and effective field theory ceases to describe the QMC data, as expected. 
%Finally, to extrapolate down to temperature regime beyond the scope of the QMC data simulations, we only use the Landau damping asymptotics $d_0^{-1} (\omega_m-\omega_n)^2+\alpha |\omega_m-\omega_n|$ to calculate the Matsubara sum. 
%This approximation yields reasonable accuracy already at temperatures $T=1/40$.\bof{Add reference to figure}. Furthermore, we use the scaling $M^2(T) = a T \log(b T)$ from above to calculate the thermal gap parameter required as input for the extrapolation.

\section{Transport properties} 
In this paragraph, we supply information about the application of the Meir-Wingreen formalism~\cite{meir1992}, originally developed for interacting mesoscopic systems, to the Kondo heterostructure. First, we derive the expression for the conductivity, Eq.~\eqref{eq:defSigma} in the main text. We then compute the retarded marginal-Fermi-liquid self-energy $\Sigma_{\psi\psi}^R(\omega, \ve k_2)$ as function of real frequencies $\omega$. This allows us to obtain the temperature dependence of the conductivity.

\subsection{Meier-Wingreen formalism}
In a situation where the noninteracting region above the spin layer is kept at a constant spatially homogeneous electric potential $e V>0$, a stationary current will flow through the spin layer without generating currents parallel to the latter. The corresponding operator for the current density in the out-of-plane direction reads
\begin{align}\label{eq:resCurMom}
\begin{split}
 \hat \jmath^z_{\ve i,R_z \to R_z+1} & = -i e t \sum_{\sigma}\left( \hat c^\dagger_{\ve i,R_z+1,\sigma} \hat c_{\ve i,R_z,\sigma} - \text{h.c.}\right)\, .
\end{split}
 \end{align}
This expression can be derived, in analogy to Ref.~\cite{caroli1971}, from the discrete form of the continuity equation $\partial_t \hat\rho_{\ve i, R_z} = -(\hat \jmath^z_{\ve i,R_z \to R_z+1} - \hat\jmath^z_{\ve i,R_z-1 \to R_z})$ for the electronic charge density  
$\hat \rho_{\ve i,R_z} = -e \sum_{\sigma} \hat c^\dagger_{\ve i,R_z,\sigma} \hat c^{}_{\ve i,R_z,\sigma}$. The above form then follows by calculating $\partial_t \rho $ via the Heisenberg equations of motions generated by the Hamiltonian, Eq.~\eqref{eq:model} of the main text. Note that the resulting $\hat\jmath^z$ contains only hopping of conduction electrons between lattice sites but no interaction terms. This holds even at $R_z=0$ since local spin flips only change the occupations numbers of the individual spin components but conserve the total particle number. Without in-plane modulations of the current-density, the total current becomes
$\hat I^z = \sum_{\ve i} \hat \jmath^z_{\ve i,R_z \to R_z+1} = L \hat \jmath^z_{\ve q_2=0, R_z \to R_z+1}$, which is independent of the layer index in a stationary situation. Consequently, we can calculate its expectation value as $\<\hat I^z\> = 1/2 \cdot L \< \hat \jmath^z_{\ve q_2=0, R_z=-1 \to 0} + \hat \jmath^z_{\ve q_2=0, R_z=0 \to 1}\>$ which corresponds, in the mesoscopic language of Ref.~\cite{meir1992}, to the averaged current from the first noninteracting lead into the interacting region and from the latter into the second noninteracting lead. Inserting Eq.~\eqref{eq:resCurMom}, we get
\begin{align}
\begin{split}
 \<\hat I^z\> = - \frac{i e t}{2L^{1/2}} \sum_{\ve k_2,k_z,\sigma}&  \left(e^{- i k_z} \<\hat c^\dagger_{\ve k_2,R_z=0,\sigma} \hat c_{\ve k_2,k_z,\sigma}\> \right. \\ 
 &\left.-
 e^{ i k_z} \< \hat c^\dagger_{\ve k_2,k_z,\sigma} \hat c_{\ve k_2,R_z=0,\sigma}\>\right) + \text{h.c} \, ,
\end{split}
\end{align} 
where the Fourier transform in the out-of-plane direction treats the two regions, with $R_z<0$ and $R_z>0$, as two identical nearest-neighbor tight-binding bands, each hosting $L$ Bloch states $|k_z\>$. To make contact with Ref.~\cite{meir1992}, we introduce the matrix elements $V=-t\exp(i k_z)/L^{1/2}$ and, moreover, the real-time propagators $i \left \langle \hat{c}^\dagger_{\ve k_2,R_z=0,\sigma} \hat c_{\ve k_2,k_z,\sigma}(t) \right \rangle = G^<_{k_z,R_z=0,\sigma}(t,\ve k_2)$ and $i \left \langle \hat{c}^\dagger_{\ve k_2,k_z=0,\sigma} \hat c_{\ve k_2,R_z=0,\sigma}(t) \right \rangle = G^<_{R_z=0,k_z,\sigma\sigma}(t,\ve k_2,\sigma)$ that contain one operator from a lead and one from the spin layer. 
With these variables, the current becomes
\begin{align}
\begin{split}
\<\hat I^z\> = \frac{e}{2} \sum_{\ve k_2,k_z \sigma} \int & \frac{d \omega}{2\pi} 
\left(V G^<_{R_z=0,k_z,\sigma}(\omega,\ve k_2) \right. \\
& \qquad \left. - V^\ast G^<_{k_z,R_z=0,\sigma}(\omega,\ve k_2)
  \right) + \text{h.c.} \, ,
\end{split}
\end{align}
which is, apart from the additional sum over $\ve k_2$ in our case, equivalent to Eq.~(2) in Ref.~\cite{meir1992}. 
The fact that the in-plane momentum is conserved and all propagators are diagonal in $\ve k_2$ allows the application of the Meir-Wingreen approach, based on single-particle Green's functions in the Keldysh formulation, directly to $G^<_{k_z,R_z=0,\sigma}(t,\ve k_2)$, for each $\ve k_2$ independently. This results in 
\begin{align}
\begin{split}
\<\hat I^z\> = i e \sum_{\ve k_2} \int \frac{d \omega}{2\pi} &  (n^{(R_z>0)}_F(\omega)-n_F^{(R_z<0)}(\omega)) \Gamma_\sigma(\omega,\ve k_2) \\
& \times 2i \Im g^R(\omega,\ve k_2)\, ,
\end{split}
\end{align}
which represents the expression corresponding to Eq.~(6) of Ref.~\cite{meir1992}. Here, $\Gamma(\omega,\ve k_2)=2\pi \sum_{k_z} |V|^2 \delta(\omega-\epsilon_{\ve k_2}+2t \cos k_z) $ contains the density of states of the leads at given energy $\omega$ and in-plane momentum $\ve k_2$, which is the same for both noninteracting regions. $\Gamma(\omega, \ve k_2)$ can be rewritten in terms of the local free propagator $g_0$ introduced below Eq.~\eqref{eq:defTmf} in the main text, $\Gamma(\omega,\ve k_2) = 2\pi t^2 a_0(\omega ,\ve k_2)$, with the noninteracting local spectral function $a_0(\omega ,\ve k_2) = -\pi^{-1} \Im g_0(i \omega_n \to \omega + i0^+)$ in real frequencies. Similarly, interaction effects are encoded in the spectral function $a(\omega,\ve k_2) = - \pi^{-1} \Im g^R(\omega,\ve k_2)$ that is obtained from Eq.~\eqref{eq:defgfull}, again via analytic continuation $\omega_n \to \omega + i 0^+$. Note that evaluating $a(\omega,\ve k_2)$ requires knowledge of the retarded $T$ matrix $\tilde T^R(\omega,\ve k_2) = \tilde T(i \omega_n \to \omega + i 0^+,\ve k_2)$. Furthermore, there is no difference between the spin components in the absence of magnetic order such that $\<\hat I_z\>$ contains a factor of two from the spin sum.
Expanding the difference of Fermi-Dirac distributions to lowest order in $eV$ yields
\begin{align}
\<\hat I^z\> = 2 e^2 t^2 \sum_{\ve k_2} \int d \omega   (\left.-n'_F(\omega)\right|_{\mu=0}) a_0(\omega,\ve k_2) a(\omega,\ve k_2) V\,.
\end{align}
Converting the above into Ohm's law $\<\hat \jmath^z \> = \sigma E_z$, with the current density $\<\hat \jmath^z \> = \<I^z\>/L^2$ and the electric field from the voltage drop $E_z=V/(2a)$  between the two leads, which are at a distance of two lattice constants $a=1$, finally leads to Eq.~\eqref{eq:defSigma} in the main text.

\subsection{Retarded composite-fermion self-energy}
As discussed above, the computation of the conductivity relies on the dressed spectral function $a(\omega,\ve k_2) = - \pi^{-1} \Im g^R_\sigma(\omega,\ve k_2)$ defined on the real frequency axis, with the local propagator $g^R(\omega,\ve k_2) = g(i \omega_n \to \omega + i 0^+)$ from Eq.~\eqref{eq:defgfull}. To study the effects of the marginal Fermi liquid on the transport properties, we have to insert the retarded beyond-mean-field $T$ matrix $\tilde T^R(\omega,\ve k_2) = [\vmf^{-2} \omega - \tilde \epsilon_{\ve k_2}-\Sigma^R_{\psi\psi}(\omega,\ve k_2) - g^R_0(\omega,\ve k_2)]^{-1}$ into $g^R$.
The analytic continuation of the perturbative self-energy, Eq.~\eqref{eq:oneloopPsi}, reads
\begin{align}\label{eq:SigmaPsiR}
\begin{split}
&\Im \Sigma^R_{\psi \psi} (\omega_k, \ve k_2) =  \frac{4 g_{\text{eff}}^2}{\Jk^2} \int  \frac{d^2 q_2}{(2\pi)^2} \int \frac{d\omega_q}{\pi}
\Im D^R(\omega_q,\ve q_2) \\
& \quad \times \Im \tilde T_\mf^R(\omega_k-\omega_q,\ve k_2-\ve q_2) [n_B(\omega_q)+n_F(\omega_q-\omega_k)]\, ,
\end{split}
\end{align} 
with the asymptotic magnetic susceptibility in the vicinity of the ordering wavevector as $D^R(\omega,\ve q_2) = (D_0^{-1}\omega^2-c_B^2 (\ve q_2-\ve Q)^2-M^2(T) +i \alpha \omega)$. In the above, we focus on the important imaginary part that captures the decay of the excitations and consider a non-self-consistent evaluation, which does not affect the scaling with temperature, as is argued below. Furthermore, we expect from the fast convergence of the self-consistency loop on the Matsubara frequencies, with typical deviations of $\Im \Sigma_{\psi\psi}$ on the order of 10\%-20\% between the first two iterations, that a perturbative treatment provides also a good quantitative estimate. 
After shifting the momentum $\ve q_2 \to \ve q_2 + \ve Q$, we rescale $\omega = T \bar \omega$ and $c_B \ve  q_2 = \sqrt{\alpha T} \bar{\ve  q}$, such that the convolution becomes
\begin{align}
\begin{split}
&\Im \Sigma^R_{\psi \psi} (\omega_k, \ve k_2)  = \\
& \frac{4 g_{\text{eff}}^2 T}{\Jk^2 c_B^2} \int   \frac{d^2 \bar q_2}{(2\pi)^2} \int \frac{d\bar \omega_q}{\pi} 
\frac{\bar \omega_q [\bar n_B(\bar \omega_q)+\bar n_F(\bar \omega_q-\bar \omega_k)]}{(T\bar\omega_q^2 /\alpha-c_B^2 \bar q_2^2-\bar M^2)^2+ \bar\omega_q^2}  \\ 
&\qquad \quad \times\Im \tilde T^R_\mf\left(T(\bar\omega_k-\bar \omega_q),\ve k_2-\ve Q -\sqrt{\alpha T}\bar{\ve q}_2/c_B)\right)\, .
\end{split}
\end{align}  
Here, we have defined the dimensionless distribution functions 
$\bar n_{F,B}(\bar \omega)  = 1/[\exp(\bar \omega) \pm 1] $ and the dimensionless mass parameter
$\bar M^2 = M^2(T)/(\alpha T)$. 
For small temperatures, the denominator of the magnetic susceptibility implies the scaling $\bar q_2 \sim \max(\bar M, \bar \omega_q^{1/2})$ while its numerator times the combination of distribution functions makes the internal frequency vary on the scale $\bar \omega_q \sim \max(1,\bar \omega_k)$. Furthermore, we are only interested in $|\omega_k|\ll t $ and simultaneously $T \ll t$ where the hopping amplitude $t$ corresponds to the energy scale of the local bare propagator $g^R_{0}$. Hence, $\sqrt{\alpha T} \ve  q_2/c_B$ is only a negligible correction to $\ve  k_2+\ve  Q$. As a result, we obtain for the leading behavior in the limit $T \to 0$
\begin{align}
\begin{split}
\Im \Sigma^R_{\psi \psi} (\omega_k,\ve k_2) & = \frac{4g_{\text{eff}}^2 T}{\Jk^2 c_B^2} \Im \tilde T^R(0,\ve k_2-\ve  Q) \int  \frac{d^2 \bar q_2}{(2\pi)^2} \\ 
& \times \int \frac{d\bar \omega_q}{\pi}
\frac{\bar \omega_q [\bar n_B(\bar \omega_q)+\bar n_F(\bar \omega_q-\bar \omega_k)]}{(\bar q_2^2+\bar M^2)^2+ \bar\omega_q^2}  \, .
\end{split}
\end{align} 
Next, we solve the momentum integral, which is UV convergent,
\begin{align}
\begin{split}
& \Im \Sigma^R_{\psi \psi} (\omega_k, \ve k_2) = \frac{2g_{\text{eff}}^2 T}{\pi \Jk^2 c_B^2} \Im \tilde T^R(0,\ve k_2-\ve Q) \\
& \times \int \frac{d\bar \omega_q}{\pi}
\frac{1}{2}\arctan\left(\frac{\bar \omega_q}{\bar M^2}\right) [\bar n_B(\bar \omega_q)+\bar n_F(\bar \omega_q-\bar \omega_k)] \, .
\end{split}
\end{align} 
Note that we do not impose a cutoff $q_2 \leq \Lambda$ since for the variable $\bar q$ the corresponding cutoff behaves as $c_B\Lambda/(\alpha T)^{1/2} \to \infty$ anyway. The last equation may be cast into a dimensionless scaling function
\begin{align}\label{eq:defSigmaRscal}
\Im \Sigma^R_{\psi \psi} (\omega_k, \ve k_2) = \frac{2g_{\text{eff}}^2 T}{\pi \Jk^2 c_B^2} \Im \tilde T^R(0,\ve k_2-\ve Q)  \hat\Sigma_{\psi\psi}^R(\bar \omega_k, \bar M)\, ,
\end{align}
which is useful for the calculation of the conductivity and, moreover, allows to access several limiting cases analytically.

For $\omega_k \to 0 $, the frequency integration is focused to the regime $|\bar \omega_q| \leq 1$ whereas larger frequencies are exponentially suppressed. Since we expect $\bar M^2 \sim 1/\log T \gg 1$, due to Hertz-Millis scaling at small $T$, we expand $\arctan(\bar \omega_q/\bar M^2) \simeq \bar \omega_q/\bar M^2$ and obtain 
\begin{align}\label{eq:scalResw0}
\hat\Sigma_{\psi\psi}^R(0, \bar M) \simeq \int \frac{d\bar \omega_q}{2\pi}
\frac{\bar \omega_q}{\bar M^2} [\bar n_B(\bar \omega_q)+\bar n_F(\bar \omega_q)] = \frac{\pi}{4 \bar M^2}  \, .
\end{align} 
Restoring units leads to the temperature dependence $\Im \Sigma^R_{\psi \psi} (\omega_k \to 0,\ve  k_2) \sim T/\log T$. If we instead tune the system away from the quantum critical regime, by replacing $M^2(T)$ with a constant value $M^2$, we find the same result for $\hat\Sigma_{\psi\psi}^R(0, \bar M)$, but obtain the Fermi-liquid behavior $\Im \Sigma^R_{\psi \psi} (\omega_k \to 0,\ve k_2) \sim T^2/M^2$.

\begin{figure}[tb]
\begin{center}
\includegraphics[width=\columnwidth]{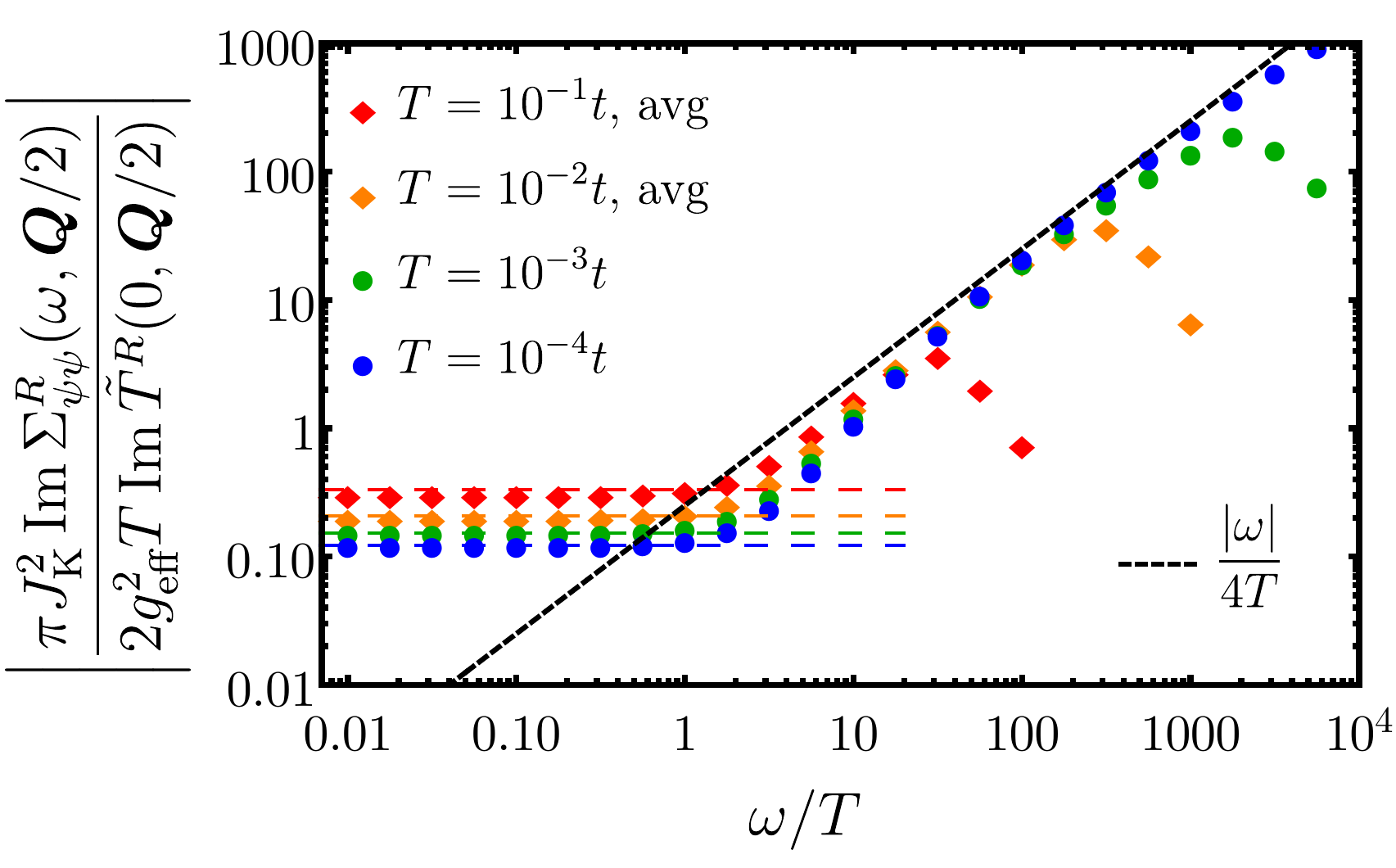}
\caption{Retarded self-energy $\Im \Sigma^R_{\psi\psi}$ in units of $T \Im \tilde T^{R}$ as function of $\omega/T$ for fixed in-plane momentum $\ve q_2 = \ve Q/2$, from effective field theory, illustrating the scaling behavior advertised in Eq.~\eqref{eq:defSigmaRscal}.
% reveals the scaling behavior shown in Fig.~\ref{fig:QMCmatch}(e).
Dashed horizontal lines correspond to values of the scaling function $\hat{\Sigma}(0, M^2(T)/(\alpha T))$. Scaling breaks down at large frequencies $\omega \sim t$ on the order of the hopping or above $T \sim 10^{-2} t$ when the low-frequency limit is no longer given by $\hat{\Sigma}(0, M^2(T)/(\alpha T))$ (not shown). As can be seen in the figure, however, averaging $\Im \Sigma^R_{\psi\psi}(\omega,\ve k_2)$ over the projected 2D Fermi surface (labeled as avg) yields good agreement with the scaling form. This explains why the comparison of the conductivity between effective field theory and QMC still shows quantitative agreement in this temperature range.}
\label{fig:scalPlot2}
\end{center}
\end{figure}

On the other hand, for frequencies $|\omega_k|\gg T$, the integral is dominated by frequencies $\bar \omega_q \sim \bar \omega_k$. Thus, we may replace the distribution functions by their zero-temperature forms. However, two scenarios have to be distinguished: (i) $\bar M^2 \ll |\bar \omega_q|$, which allows to expand $\arctan(\bar \omega_q/\bar M^2) \to \pi/2 \cdot \text{sgn}(\bar \omega_q)$ and (ii) $\bar M^2 \gg |\bar \omega_q|$ such that $\arctan(\bar \omega_q/\bar M^2) \to \bar \omega_q / \bar M^2$ again. In the first case, we recover the marginal-Fermi-liquid scaling known from the zero-temperature analysis,
\begin{align}\label{eq:scalReswMFL}
\begin{split}
& \hat \Sigma_{\psi\psi}^R(|\bar \omega_k| \gg 1, \bar M \ll |\bar \omega_k|^{1/2}) \simeq \\
&\qquad \int \frac{d\bar \omega_q}{4} \text{sgn}(\bar \omega_q) [\theta(\bar \omega_k-\bar \omega_q)-\theta(-\bar \omega_q)] = \frac{1}{4}|\bar \omega_k| \, .
\end{split}
\end{align} 
In contrast, in the second case, we find Fermi-liquid behavior, as expected for the ground state of the paramagnetic heavy-fermion metal,
\begin{align}\label{eq:scalReswFL}
\begin{split}
&\hat \Sigma_{\psi\psi}^R(|\bar \omega_k| \gg 1, \bar M \gg |\omega_k|^{1/2}) \\
& \simeq \int \frac{d\bar \omega_q}{2\pi} \frac{\bar\omega_q}{\bar M^2}(\bar \omega_Q) [\theta(\bar \omega_k-\bar \omega_q)-\theta(-\bar \omega_q)] = \frac{\bar \omega_k^2}{4 \pi \bar M^2} \, ,
\end{split}
\end{align}
which is equivalent to $\Im \Sigma^R_{\psi\psi}(\omega, \ve k_2) \sim \omega^2$, since we formally tune the system into the Fermi-liquid phase. In sum, we find in the paramagnetic heavy-fermion-metal phase $|\Im \Sigma^R_{\psi\psi}(\omega\ve k_2)| \sim \max(\omega^2,T^2)$. 
In the quantum critical regime, however, the Fermi-liquid asymptotics is constrained to the regime $M^2(T) \sim T \log T \gg |\omega| \gg T $, while already at frequencies of order $T \log T$ the crossover to the marginal-Fermi-liquid behavior sets in, such that the Fermi-liquid regime is essentially completely hidden. This is confirmed by the numerical evaluation of Eq.~\eqref{eq:defSigmaRscal}, using the numerical parameters from the matching procedure outlined in the previous section. As is shown in Fig.~\ref{fig:QMCmatch}(e) in the main text, no intermediate Fermi-liquid regime emerges in the quantum critical regime. In total, we can summarize the quantum critical behavior of the retarded single-fermion self-energy as
\begin{align}\label{eq:SigmaRasy}
\begin{split}
\Im \Sigma^R_{\psi \psi} (\omega_k,\ve k_2) \simeq & \frac{g_{\text{eff}}^2}{2\pi \Jk^2 c_B^2} \Im\, T^R_\mf(0,\ve k_2-\ve Q) \\
& \times
\begin{cases}
\dfrac{\pi \alpha T}{a \log(b T)} \, & \text{ if } \bar \omega_k \to 0\,, \\
|\omega_k| \, & \text{ if } |\bar \omega_k| \gg \bar M^2 \gg 1 \,,
\end{cases}
\end{split}
\end{align}
where we have inserted the parametrization $M^2(T) = a T \log(b T)$ introduced above.
Figure~\ref{fig:scalPlot2} confirms that $\Im \Sigma^R$ indeed obeys the above scaling behavior, which is also depicted in Fig.~\ref{fig:QMCmatch}(e).
Finally, we note that self-consistent corrections cannot change the temperature dependence of the perturbative results. First, one repeats the calculation with the dressed $T$ matrix $\tilde T^R(\omega,\ve k_2) = [\vmf^{-2} \omega - \tilde \epsilon_{\ve k_2}-\Sigma^R_{\psi\psi}(\omega,\ve k_2) - g^R_0(\omega,\ve k_2)]^{-1}$ inserted into Eq.~\eqref{eq:SigmaPsiR}. After introducing the dimensionless variables $\bar \omega_q$ and $\bar{\ve {q}}_2$ again, $\Im\Sigma^R_{\psi\psi}$ is seen to scale away in the limit $T \to 0$ for all external frequencies $|\omega_k| \ll t$, while the real part merely gives rise to a renormalization of mean-field dispersion $\tilde\epsilon_{\ve k_2}$. Consequently, the results from above still apply with the replacement $\tilde T^R_\mf(0,\ve k_2-\ve Q) \to \tilde T^R(0,\ve k_2-\ve Q)$. As already mentioned above, the quantitative effect is expected to be small. 

\subsection{Conductivity from effective field theory}
\begin{figure}[t!]
%\begin{center}
\includegraphics[width=\columnwidth]{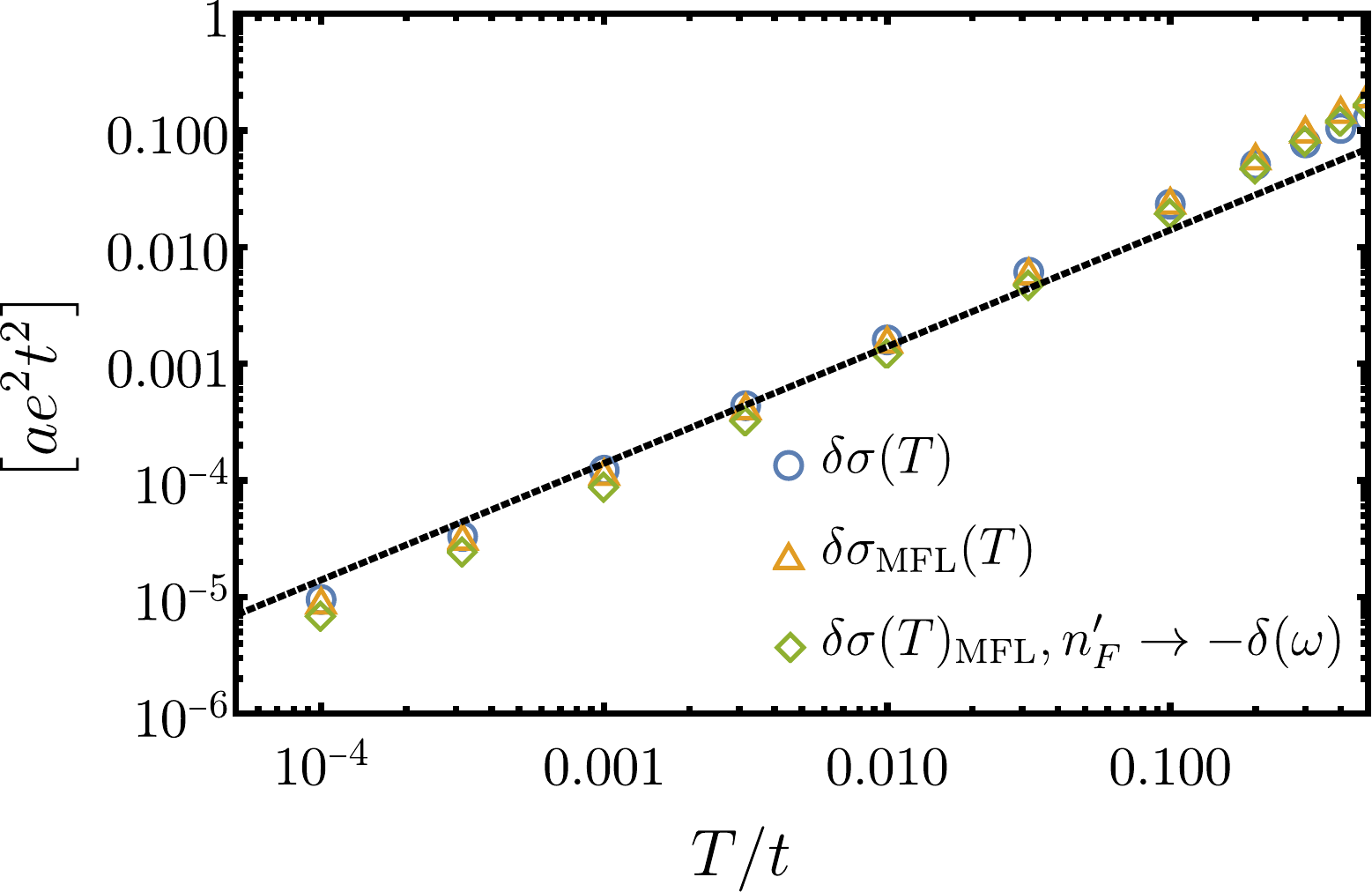}
\caption{
Conductivity as function of temperature from effective field theory in the quantum critical regime at different levels of approximations. 
Blue: Difference $\delta \sigma(T)=\sigma(T)-\sigma(0)$ with the excat conductivity from Eq.~\eqref{eq:defSigma}. Yellow: Approximate result $\delta \sigma_{\text{MFL}}(T)$ from Eq.~\eqref{eq:deltasT}, focussing on the marginal-Fermi-liquid excitations. 
Green: $\delta \sigma_{\text{MFL}}(T)$ with the additional replacement $n'_F(\omega) \to -\delta( \omega)$. The dashed black line shows the expected $T$ linear variation.}
%Blue: Deviation of the exact deviation~\eqref{eq:defDeltaSigma} %\lj{add correct equation label}
%for of conductivity $\sigma(T)$ from its value at $\sigma(T=0)$ at $\Jkc$. Yellow: Approximate result $\delta \sigma(T)$ from Eq.~\eqref{eq:deltasT} focused on the marginal-Fermi-liquid excitations. Green: $\delta \sigma(T)$ with the additional replacement $\bar n'_F(\bar \omega) \to -\delta(\bar \omega)$. The dashed black line shows the expected $T$ linear variation.
\label{fig:dsigma}
%\end{center}
\end{figure}

Finally, we evaluate the temperature dependence of the conductivity $\sigma(T)$ from Eq.~\eqref{eq:defSigma} in the main text, using again units in which $a=1$. At $T=0$, we can replace $n'_F(\omega)|_{\mu=0} = -\delta(\omega)$ and obtain
\begin{align}\label{eq:resSigmaGS1}
\sigma(T=0) =  4 \pi e^2 t^2 \int \frac{d^2 k_2}{(2\pi)^2}  a_0(0, \ve k_2) a(0,\ve k_2) \, ,
\end{align}
which yields, with the parameters from the matching procedure, $\sigma(T=0) \approx 0.0051$. Note that this value is not affected by the presence or absence of marginal-Fermi-liquid excitation since $\Im \Sigma^R_{\psi\psi}|_{T=0} \sim |\omega|$ drops out from the dressed spectral function due to the factor $\delta(\omega)$. Physically, only the conventional, mean-field-like scattering processes between conduction electrons and the Kondo-screened spins contribute to $\sigma(T=0)$. At finite temperatures, however, both the effective field theory and QMC simulations observe a linear growth of $\sigma(T)$ in the quantum critical regime, see Figs.~\ref{fig:sigma} and \ref{fig:dsigma}.  
In contrast to the behavior in the limit $T \to 0$, the latter temperature dependence is a direct consequence of the marginal Fermi liquid. Similarly, the quadratic increase of $\sigma(T)$ in the paramagnetic heavy-fermion metal can be attributed to the Fermi-liquid excitations. This can be best understood in terms of the deviation $\delta \sigma(T) = \sigma(T) -\sigma(T=0)$. The lowest-order correction in the beyond-mean-field decay rate to the dressed spectral function reads $a_1(\omega,\ve k_2) -\pi^{-1} \Re[g_0(\omega,\ve k_2)^2 \tilde T_\mf^R(\omega,\ve k_2)^2] \Im \Sigma^R_{\psi\psi}(\omega,\ve k_2) $, cf.\ Eq.~\eqref{eq:defgfull}. Again, we neglect $\Re \Sigma^R_{\psi\psi}$, which is expected to provide merely a small renormalization of $\tilde \epsilon_{\ve k_2}$, and obtain
\begin{align}\label{eq:defDeltaSigma}
\begin{split}
& \delta\sigma_{\text{MFL}}(T) = 4    e^2 t^2 \int \frac{d^2 k_2}{(2\pi)^2} d\omega\,  a_0(\omega, \ve k_2)
\left. n_F'(\omega)\right|_{\mu=0} \\
&\qquad\qquad \times \Re[g_0(\omega,\ve k_2)^2 \tilde T_\mf^R(\omega,\ve k_2)^2] \Im \Sigma^R_{\psi\psi}(\omega,\ve k_2) \, .
\end{split}
\end{align}
Inserting now the scaling form, Eq.~\eqref{eq:defSigmaRscal}, and rescaling $\bar{\omega} = \omega/T$, yields
in the limit $T \to 0$
\begin{align}
\label{eq:deltasT}
\begin{split}
&\delta\sigma_{\text{MFL}}(T) = \frac{8g_{\text{eff}}^2 e^2 t^2 T}{\pi \Jk^2 c_B^2} \int \frac{d^2 k_2}{(2\pi)^2} d\bar \omega\,  a_0(0, \ve k_2) \bar n_F'(\bar \omega) \\
&\! \times \!\Re\![g_0(0,\ve k_2)^2 \tilde T_\mf^R(0,\ve k_2)^2]\Im \tilde T_\mf^R(0,\ve k_2-\ve Q)  \hat\Sigma_{\psi\psi}^R(\bar \omega_k, \bar M) ,
\end{split}
\end{align}
with $\bar M^2 = M^2(T)/(\alpha T)$.
First, we observe that $\delta\sigma_{\text{MFL}}(T)$ is positive: The mean-field $T$ matrix at zero frequency is dominated by the bare local propagator $\tilde T_\mf^R(0,\ve k_2-\ve Q) \simeq -1/g_0(0,\ve k_2-\ve Q)$. Furthermore,
$\Im \tilde T_\mf^R(0,\ve k_2-\ve Q) $ is only finite in the projected 2D Brillouin zone where $g_0(0,\ve k_2-\ve Q)$ is purely imaginary and $\Im g_0(0,\ve k_2-\ve Q) < 0$. Together with $\bar n_F(\bar \omega) <0$ and $\hat\Sigma^R_{\psi\psi} >0$ one concludes then $\delta\sigma_{\text{MFL}}(T) >0$. The growth of the conductivity with increasing temperature reflects the reduction of Kondo screening by thermal fluctuations.
Moreover, since $\hat\Sigma_{\psi\psi}^R(\bar \omega_k, \bar M)$ contributes only logarithmic corrections, according to Eqs.~\eqref{eq:scalResw0} and~\eqref{eq:scalReswMFL}, we indeed find $\delta\sigma_{\text{MFL}}(T) \sim T$ in the quantum critical regime. This is confirmed in Fig.~\ref{fig:dsigma}, which shows that the expansion in Eq.~\eqref{eq:deltasT} properly captures the behavior of the exact $\delta \sigma(T)=\sigma(T)-\sigma(0)$, with the full Meir-Wingreen conductivity from Eq.~\eqref{eq:defSigma}. In the paramagnetic heavy-fermion metal, we replace $M^2(T)$ by a temperature-independent constant $M^2$, which leads, via the results of Eqs.~\eqref{eq:scalResw0} and~\eqref{eq:scalReswFL}, to a Fermi-liquid self-energy $\Im \Sigma^R_{\psi \psi} \sim \max(T^2,\omega^2)$. Replacing the MFL self-energy by the latter FL form in Eq.~\eqref{eq:defDeltaSigma}, implies the scaling $\delta \sigma_{\text{FL}}(T) \sim T^2$, which is also observed in the QMC simulations in Fig.~\ref{fig:sigma} in the main text.   
We also note that approximating $\bar n'_F(\bar \omega) \to \delta(\bar \omega)$ at finite temperatures does not change the scaling of $\delta\sigma(T)_{\text{MFL}}$, but merely replaces $\hat \Sigma(\bar \omega,\bar M) \to \hat \Sigma(0,\bar M)$. As is shown in Fig.~\ref{fig:dsigma}, this approximation, used to calculate the conductivity from the QMC data, see below, only introduces a deviation on the order of 20\%.  

Finally, we mention that the effective one-dimensional van-Hove singularities of $a_0(\omega,\ve k_2)=-\pi^{-1} \Im (\sqrt{\omega+i 0^+ - \epsilon_{\ve k_2} +2t}\sqrt{\omega+i 0^+ - \epsilon_{\ve k_2} -2t})^{-1}$ at the boundaries of the projected 2D Fermi surface, see Fig.~\ref{fig:QMCmatch}(b) in the main text, where $\epsilon_{\ve k_2} = \pm 2t$, only give rise to subleading corrections. In the vicinity of the van-Hove singularities, the dressed spectral function becomes $a(\omega,\ve k_2) \to \pi^{-1} \Im \Sigma^R_{\psi\psi}(\omega,\ve k_2)$. Parametrizing the 2D momentum integral as a contour integral $\ve k_2(l)$ along the singular lines in the Brillouin zone and a momentum component $k_\perp$ perpendicular to the curve $\ve k_2(l)$, we obtain
\begin{align}
\delta \sigma_{\text{vH}}(T) \sim \int_0^1 dl \int \frac{dk_\perp}{(2\pi)^2} \int d\bar \omega 
\frac{T \hat \Sigma^R(\bar \omega,\bar M)}{\sqrt{T \bar \omega - v_{F}(\ve k_2(t))k_\perp }}  \, ,
\end{align}  
where $v_F(\ve k_2)$ refers to the in-plane Fermi velocity. Rescaling $k_\perp$ by $T$ leads then to $\delta \sigma_{\text{vH}}(T)\sim T^{3/2}$ for the marginal-Fermi-liquid case and to $\delta \sigma_{\text{vH}}(T)\sim T^{5/2}$ for the Fermi liquid. 

\subsection{Conductivity from QMC}
\begin{figure}[t]
%\begin{center}
\includegraphics[width=\columnwidth]{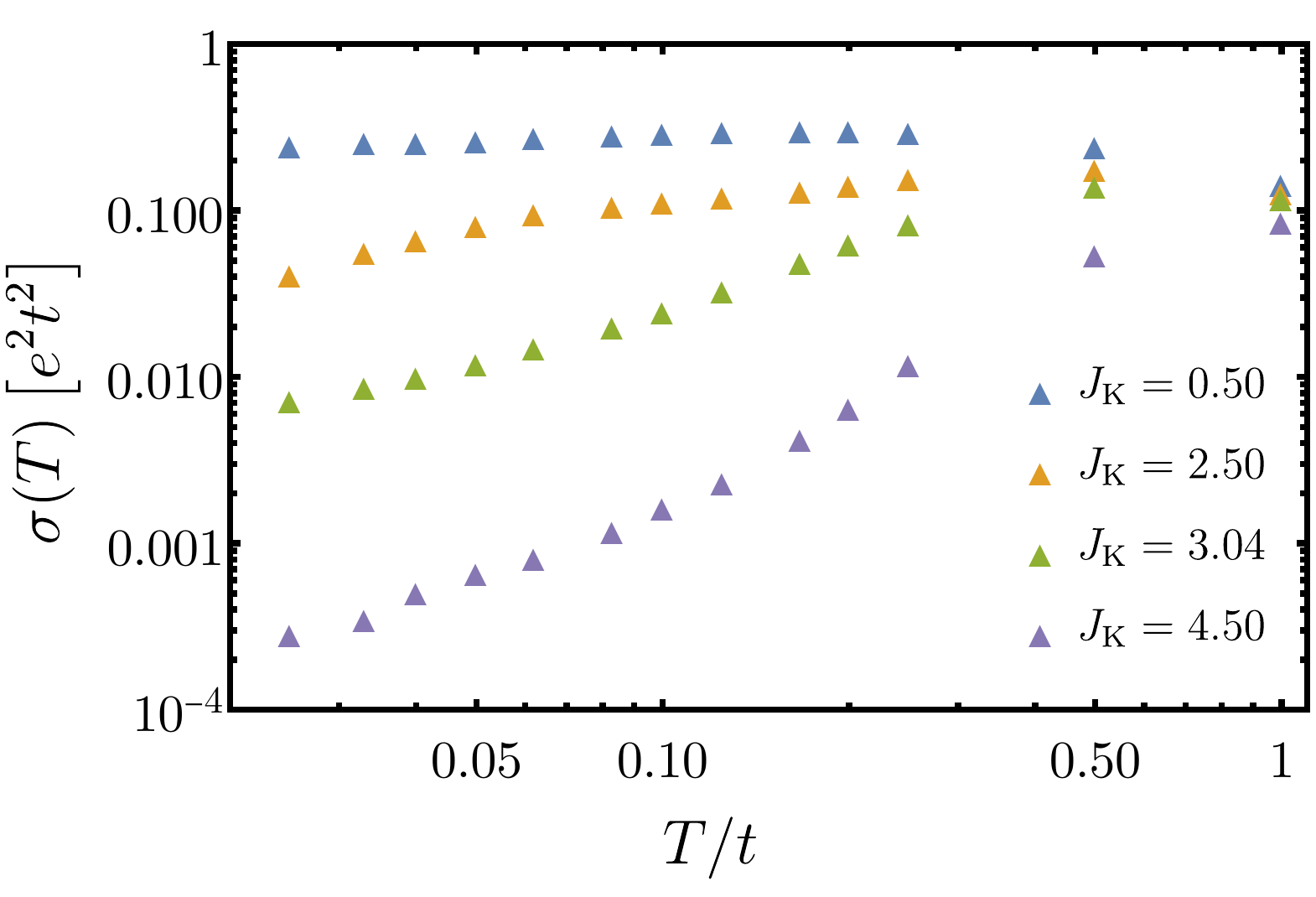}
\caption{Conductivity as function of temperature from QMC simulations for different values of $\Jk$ in the antiferromagnetic heavy-fermion phase ($\Jk = 0.5$ and $\Jk=2.5$), in the quantum critical regime ($\Jk=3.04$), and the paramagnetic heavy-fermion phase ($\Jk=4.5$).}
\label{fig:sigmaQMC}
%\end{center}
\end{figure}

The exact expression for the conductivity, Eq.~(\ref{eq:defSigma}) in the main text, requires information about the dressed Green's function at real frequency. Due to the fact that the QMC simulation are implemented in imaginary time, the real-frequency Green's function is typically obtained from numerical analytic continuation methods, which often introduce ambiguities~\cite{beach2004identifying}. To circumvent this problem, in this work, we introduce instead an approximative formula, based on Eq.~(\ref{eq:defSigma}), and express the conductivity as function of the imaginary-time Green's function. The approximation is obtained as follows: First, we replace $n'_F(\omega)|_{\mu=0} = -\delta(\omega)$ at sufficiently low temperature, as previously discussed, such that the conductivity becomes formally equivalent to Eq.~(\ref{eq:resSigmaGS1}). Furthermore, we estimate the spectral function at zero frequency by using the relation  $a(\omega=0,k)\approx\lim_{\beta\rightarrow\infty}\frac{\beta}{\pi}g(\tau=\frac{\beta}{2},\boldsymbol{k}_2)$. As a result, the conductivity can be computed by substituting the latter relation into Eq.~(\ref{eq:resSigmaGS1}), 
\begin{equation}
\sigma(T)=\pi e^{2}\frac{2\beta^{2}}{\pi^{2}}\frac{1}{L^{2}}\sum_{k_2}g_{0}\left(\tau=\frac{\beta}{2},\boldsymbol{k}_{2}\right)g\left(\tau=\frac{\beta}{2},\boldsymbol{k}_{2}\right)\, ,\label{eq:conduct_qmc}
\end{equation}
where the sum of the spin index generates a factor of two. We also replace the momentum integral by the discrete momentum summation in the finite-size lattice underlying the numerics. The quantity $g_{0}(\tau=\frac{\beta}{2})$ is the free-fermion Green's function inside the lead. All the quantities on the right-hand side of Eq.~(\ref{eq:conduct_qmc}) are accessible in the QMC calculations without analytic continuation. Hence, we expect that Eq.~(\ref{eq:conduct_qmc}) offers a much more reliable way to compute the conductivity in our model, provided that we consider low enough temperatures. Results of this evaluation are presented in Fig.~\ref{fig:sigmaQMC}, which includes, in addition to the data shown already in Fig.~\ref{fig:sigma} in the main text, also data in the antiferromagnetic heavy-fermion phase.

%%%%%%%%%%%%%%%%%%%%%%%%%%%%%%%%%%%%%%%%%%%%%%%%%%%%%%%%%%%%%%%%%%%%%%%
% ADD CLEARPAGE TO CIRCUMVENT BUG IN REVTEX
\clearpage
%%%%%%%%%%%%%%%%%%%%%%%%%%%%%%%%%%%%%%%%%%%%%%%%%%%%%%%%%%%%%%%%%%%%%%%

\end{document}